\documentclass[]{spie}  

\DeclareUnicodeCharacter{2212}{-} 
 
\usepackage{amsmath,amsfonts,amssymb}
\usepackage{graphicx}
\usepackage[colorlinks=true, allcolors=blue]{hyperref}

\usepackage{siunitx}
\newcommand{\SIadj}[2]{\SI[number-unit-product={\text{-}}]{#1}{#2}}

\title{Millimeter and sub-millimeter characterization of polymers used for infrared filters in high-sensitivity cryogenic microwave telescopes}

\author[a]{Miranda Eiben}
\author[a]{Rustam Balafendiev}
\author[a]{Thomas Gascard}
\author[a,b]{Jon E. Gudmundsson}
\affil[a]{Science Institute, University of Iceland, Sæmundargata 2, 102 Reykjavík, Iceland}
\affil[b]{The Oskar Klein Centre, Department of Physics, Stockholm University,
AlbaNova, SE-10691 Stockholm, Sweden}

\authorinfo{Further author information: (Send correspondence to M.E.)\\M.E.: E-mail: miranda@hi.is}

\begin{document} 
\maketitle

\begin{abstract}
Vacuum windows and infrared filters are important transmissive optical components in millimeter receivers, as they hold out the atmosphere and reduce radiative loading on cold stages, thereby improving cryogenic performance. However, the complex optical properties of the materials commonly used for windows and filters are poorly characterized, particularly in-band and in the sub-millimeter regime. The absorption and scattering properties of these materials are becoming increasingly important for designing high-sensitivity millimeter instruments, as their loosely constrained properties are one of the greatest sources of uncertainty remaining in noise modeling. We report the absorption in the millimeter and sub-millimeter regime of nylon 6, nylon 6/6, PTFE and polyethylene (both bulk HDPE and foam HDPE used in radio transparent filter stacks). Additionally, we report the relative power scattered out of the main beam by these materials from 90 to 330 GHz, measured in free space by a robot-enabled scanning vector network analyzer. 
\end{abstract}


\section{Introduction}
\label{sec:intro}  

The millimeter and sub-millimeter regime is rich with interesting astrophysical sources, including (but not limited to) galactic star formation, accretion disks around black holes, extra-galactic studies of high energy electrons with the Sunyaev–Zeldovich effect, and the oldest light in the universe, the Cosmic Microwave Background (CMB). However, the relatively long wavelength in this regime is in the Rayleigh tail of all blackbody sources hotter than 3 K; this complicates astrophysical ground-based measurements due to the high ambient temperatures on Earth.

Therefore, all millimeter or sub-millimeter instruments are cryogenic to limit in-band instrumental contributions. Optics are cooled to 4 K or below, where possible. Infrared radiative loading on those cold optics and on sub-kelvin detectors must be limited with transmissive infrared (IR) filters \cite{Keck2015B,bicep3}. The published optical properties of materials used for these filters, however, is limited and occasionally contradictory \cite{Lamb1996,BIRCH1981,Afsar1987,Halpern1986}. The published scattering properties of these materials are essentially non-existent, though may be as equally important as a material's absorption coefficient \cite{CorriganLiam2019Doam}. 

Excess scattered power leaving the window of a cryostat is typically expected to terminate on absorptive forebaffles attached to the front of the receiver. These forebaffles are at the ambient temperature, however, so even small amounts of scattered power can result in relatively large amounts of instrument loading. In BICEP3, for example, forebaffle on/off measurements have indicated that approximately \SI{0.34}{\pico\watt} of power on the detectors is sourced from forebaffle loading, which is approximately a third of the estimated total in-band instrument loading \cite{EibenThesis}. 

Identifying optical sources of this scattered power may allow for significant return on instrument loading, and thus reduce instrumental white and 1/f noise. Lowering noise is directly tied to survey speed returns. The FOV of CMB `small aperture telescopes' (such as the BICEP/\emph{Keck} series of instruments and the Simons Observatory SATs) is typically around 30 degrees \cite{bicep3,SO_2024}; we can therefore assume that any excess power scattered by an optic out past 15 degrees is likely to terminate on forebaffles, particularly if the optical element is high in the optical chain.   


There are two physical mechanisms that should result in transmitted incident power scattering to high angles through a material: diffraction off variable surface structure or off variable structure within the material. Hypothetically, the internal structure of bulk polymers should be significantly smaller than a wavelength, which would put any scattering well within the Rayleigh scattering regime with negligible internal differences in index resulting in negligible scattered power. However, materials such as foam polyethylene may have internal structure a significant fraction of the wavelength, which is explored further in Section \ref{sec:meas}.

In this proceedings we discuss the material selection considerations for this initial exploration in Section \ref{sec:mat_sel}. Our measurement apparatus are described in Section \ref{sec:meas}, and the measurements (including millimeter wavelength high angle scans of scattered power and millimeter and sub-millimeter transmission measurements) are shown in Section \ref{sec:results}. Finally, we conclude with next steps in Section \ref{sec:conclusion}.

\subsection{Material Selection} \label{sec:mat_sel}

A variety of polymers and polymer foams are used for transmissive optics in mm and sub-mm instruments. We have limited our measurements to four of the most common and easily obtainable polymer sheets for the initial study reported in this proceedings. 

High density polyethylene (HDPE) is a bulk polymer commonly used for windows and lenses \cite{bicep3,delessandro2018,Nakato2024}. The optical properties of HDPE are likely very similar to other commonly used polyethylenes such as ultra high molecular weight polyethylene (UHMWPE)\cite{window_paper,delessandro2018}. It is highly transmissive at millimeter frequencies, and is used here as a baseline transmissive bulk polymer. 

Polytetrafluoroethylene (PTFE) has been used as an IR filter material for CMB receivers in the past---though its use in modern CMB instruments is rare---while other millimeter and submillimeter experiments continue to use bulk PTFE filters \cite{Keck2015B,Grimes2020a}. However, PTFE membranes are commonly used for anti-reflection layers on a variety of polymer optics \cite{Dierickx2021,Shitvov2022,Eiben2024}. We measured bulk PTFE for completeness and for future comparison to PTFE anti-reflection coats.

Nylon is a common cryogenic IR filter material for millimeter receivers \cite{bicep3,Dierickx2021,Nakato2024}. It is known to be somewhat lossy at millimeter wavelengths, but the absorption is known to significantly increase above approximately 400 GHz \cite{Lamb1996}. However, there are two common formulations of bulk nylon (referred to as nylon 6 or nylon 6/6 throughout this work) and it is uncommon to report which formulation is used for a filter. We acquired both from McMaster-Carr\footnote{Nylon 6: `Machinable Cast Nylon Sheets', \url{https://www.mcmaster.com/products/nylon/shape~sheet/machinable-cast-nylon-sheets~~/}.\\ Nylon 6/6: `Wear-Resistant Nylon Sheets', \url{https://www.mcmaster.com/products/nylon/shape~sheet/wear-resistant-nylon-sheets~~/}} to test if there are significant differences in optical properties between these formulations. However, all of our nylon 6 samples arrived with fly-cut grooves on order 10 $\mu$m deep. The significance of this surface structure is discussed further in Section \ref{sec:results}.

Polymer foams have been used as both vacuum windows and IR filters in a variety of millimeter experiments \cite{Choi_2013,Keck2015B,Goldfinger2022}. Among the most common polymer foams are made by Zotefoam\footnote{\url{https://www.zotefoams.com/our-materials/azote/plastazote/}}; two formulations are common in millimeter instruments, HD30 (made from high density polyethylene) and LD24 (made from low density polyethylene), though occasionally any formulation will be referenced only as the company name. Thick blocks of Zotefoam are typically used for vacuum windows, while stacks of thinner layers are used as radio-transparent multi-layer insulation (RT-MLI) \cite{Choi_2013,Goldfinger2022}. The incredibly low density of these foams result in very low indexes of refraction, but also produce relatively large cell sizes on order 400 $\mu$m \cite{Alex2026foam}. Recent work has shown that these foams likely scatter a significant amount of power in-band \cite{Alex2026foam}.

We expect to expand our study out to other transmissive optical elements, including alumina ceramic (commonly used as an IR filter), anti-reflection coats of a variety of styles, thin composite high modulus polyethylene windows, and other commonly used filters such as the Cardiff metal-mesh low pass edge filters in the near-future \cite{Inoue2014,window_paper,Ade2006,Nadolski2020}. We focus in this work on bulk polymers or polymer foams used in CMB experiments; there is a possibility that anti-reflection coatings may also contribute to scattered power. Either structured `meta-material' or membrane `layered' anti-reflection coats may have mechanisms that scatter power out to high angles (see the related work by co-author Rustam Balafendiev, SPIE poster 14156-193) though empirically characterizing those potential effects is beyond the scope of this proceedings.

\section{Measurements}
\label{sec:meas}

\subsection{Millimeter Scatterometry}

\begin{figure}[t]
    \centering
    \includegraphics[width=0.43\linewidth]{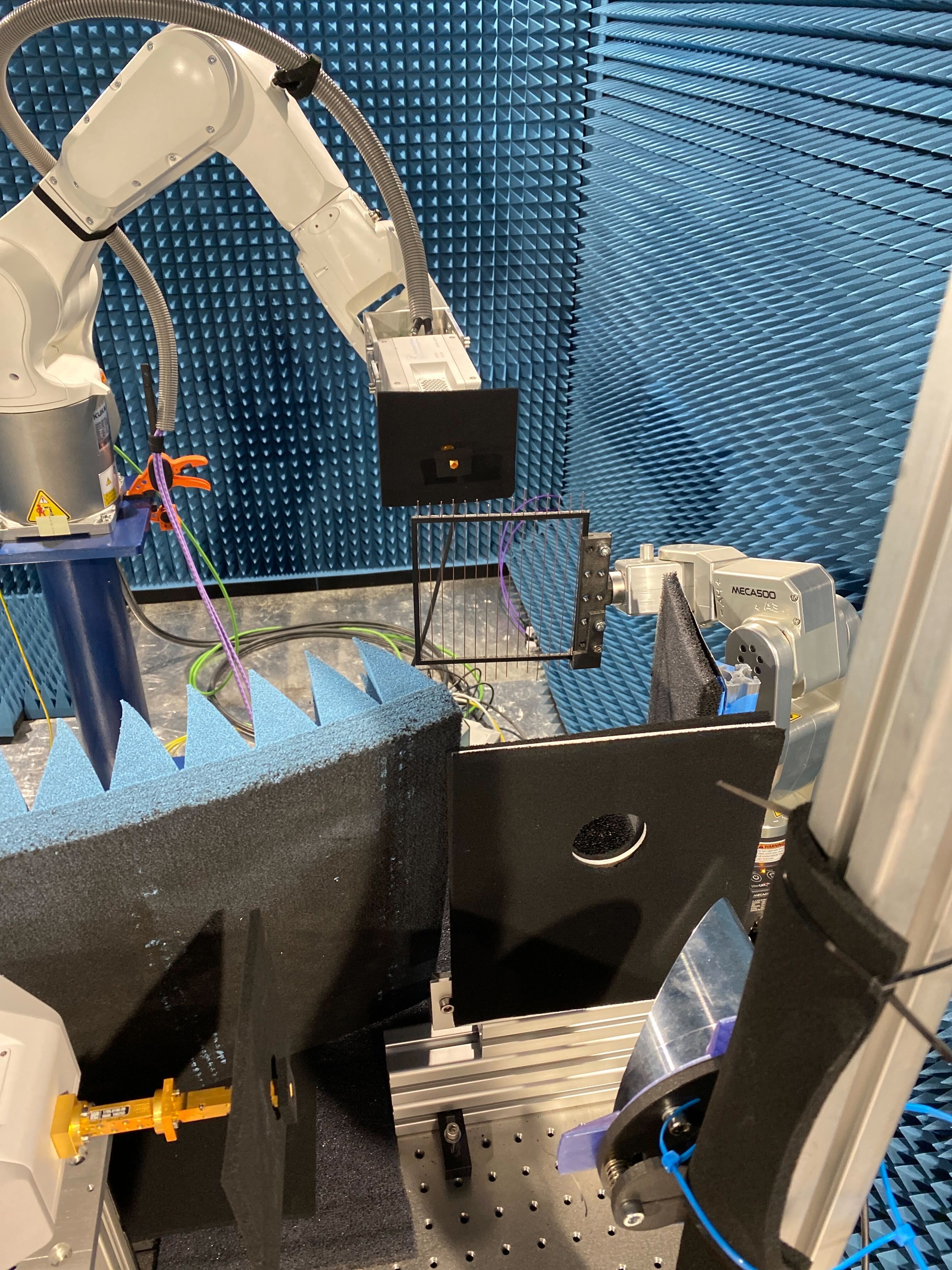}
    \includegraphics[width=0.46\linewidth]{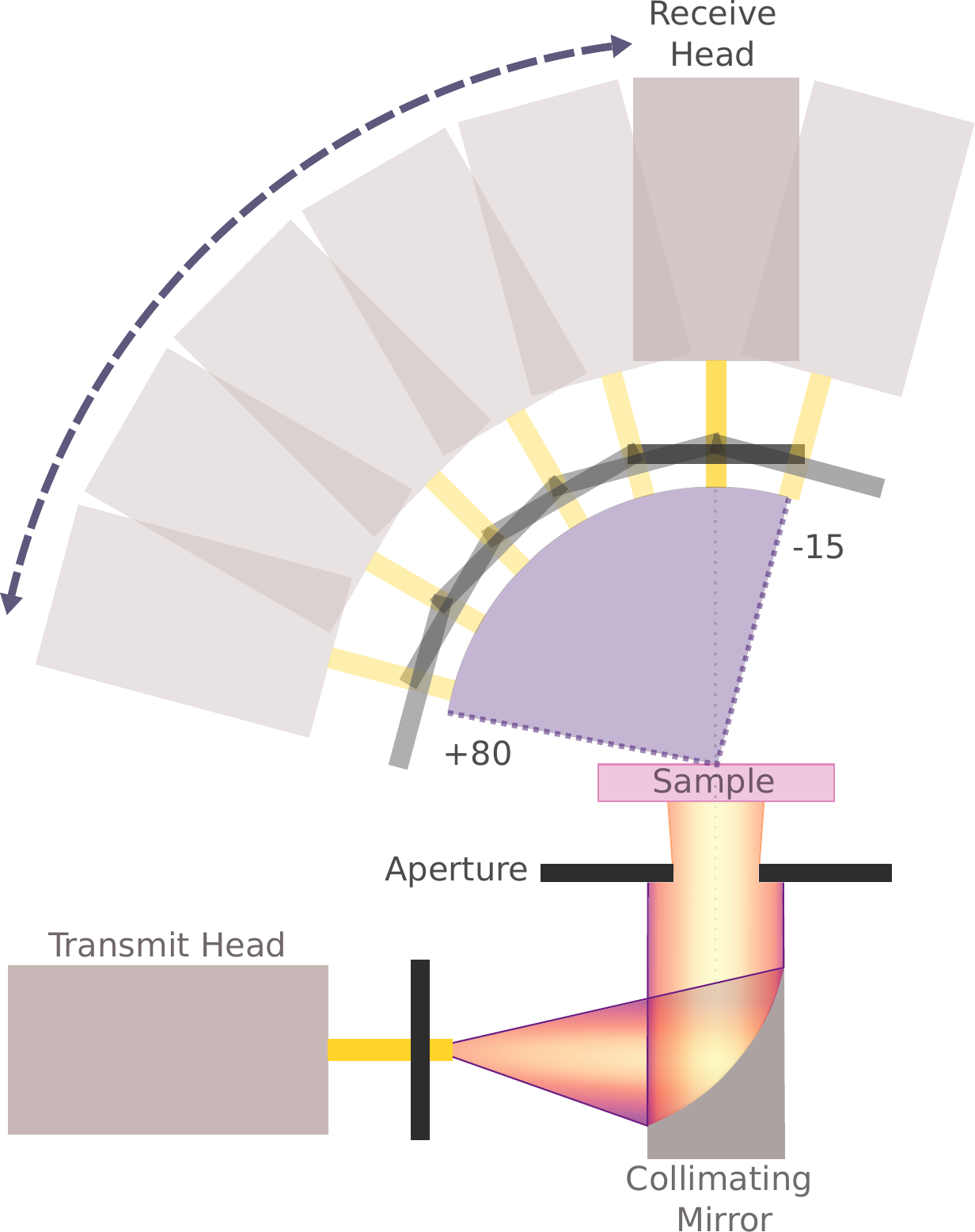}
    \caption{Scattering measurement set-up. Photo of measurement set up [left]; in the lower left is the VNA transmitter head, the lower right the collimating mirror, just above which is the \SIadj{5}{\centi\meter} diameter aperture. After the aperture is the MECA 500 sample holding robot and past that the KUKA robot holding the receiver VNA head. [Right] Diagram of the same set up, with a cartoon of the light path through the components.}
    \label{fig:scatter}
\end{figure}

We utilize a robot enabled vector network analyzer (VNA) to take measurements of transmitted scattered power out to high angles. The VNA is a Rohde and Schwarz ZNA26A with a ZC220 and a ZRX220 (J-band, 220-\SI{330}{\giga\hertz}) frequency extending transmitter and receiver head. The transmitted signal is collimated by a \SIadj{10}{\centi\meter} diameter parabolic mirror, and the beam is truncated down by a \SIadj{5}{\centi\meter} aperture plate. Samples are held in the beam by a MECA-500 6-axis robot arm, and we optimize the sample position with the small robot by maximizing the reflected (S11) signal off the sample. While this was very convenient for alignment, the MECA-500 has a very limited maximum weight it can support of 500 grams, which in turn severely limits the size of the samples it can hold. The receiver head is attached to the tool head of a KUKA KR 6 R900-2 6-axis robot, which is controlled by the custom Python-based control script described in Balafendiev et al., 2024 \cite{Balafendiev2024robot}. A photo and diagram of this set-up is shown in Figure~\ref{fig:scatter}. Collimation and focus checks are easily obtained by utilizing the robot to empirically measure phase front distributions at different transmitter positions. 

The angular extent of our scans is limited by the distance between the `wrist' joint (aka joint 5) and the base of the KUKA robot, as the receiver head (the `hand' or `end effector') must be pointed at the center of the sample. The limits of the surface that we can scan over are therefore set by the overlap of two spheres, one centered at the robot base and one centered at the sample. We offset the sample center from the robot base to maximize the angles we can achieve, at the expense of symmetric scans. 

We take two sets of reference scans: one of the empty (no sample) set-up to test minimal scattering and one with a sparse \SIadj{1}{\centi\meter} spaced wire grid to test maximal scattering. The no sample scan is used to normalize the sample measurements (discussed further in Section \ref{sec:vna}), while the wire grid is used to confirm that the scan radius is consistent with the sample center, as the angle and frequency dependence of the diffraction maxima are easily calculated from the wire spacing.

\subsection{Millimeter and Sub-millimeter Transmission}

\begin{figure}
    \centering
    \includegraphics[width=0.6\linewidth]{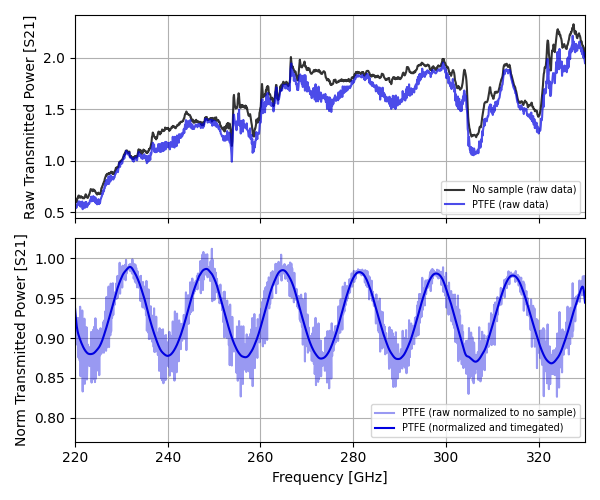}
    \caption{Example of processing of the raw VNA spectra. [Top] raw VNA spectra at the center of the beam for no sample (black) and bulk 1/4" PTFE (blue). [Bottom] The same PTFE spectra normalized by the no sample measurement (light blue) and that same normalized data timegated (dark blue).}
    \label{fig:processing}
\end{figure}

\subsubsection{Vector Network Analyzer ($220$-\SI{330}{\giga\hertz})}\label{sec:vna}

We take reference measurements to normalize the VNA transmission spectra. The normalization measurement is taken without a sample to characterize the system, then we insert the sample for the exact same measurement. The frequency spectra across the scan are then normalized to the measured spectra at the beam center of the no sample measurement, and may be further processed by timegating to a specific range. Each of these processing steps are shown in Figure \ref{fig:processing} for the PTFE sample. All scans are normalized and timegated according to this procedure.

Timegating is a process by which our VNA data is filtered in the time/spatial domain. VNA frequency spectra are Fourier transformed and then filtered to exclude cavities/resonant periods larger than a specified width. We use a Blackman-Harris filter with a gate width that excludes resonant periods longer than 3.6 nanoseconds/resonant cavities larger than 0.54 meters. As shown in Figure \ref{fig:processing}, this removes all of the high frequency noise associated with those long resonant cavities while retaining all of the relevant information pertaining to the sample measurement. 

\subsubsection{TeraScan 1550 ($70$-\SI{1400}{\giga\hertz})}\label{sec:terascan}

\begin{figure}
    \centering
    \includegraphics[width=0.7\linewidth]{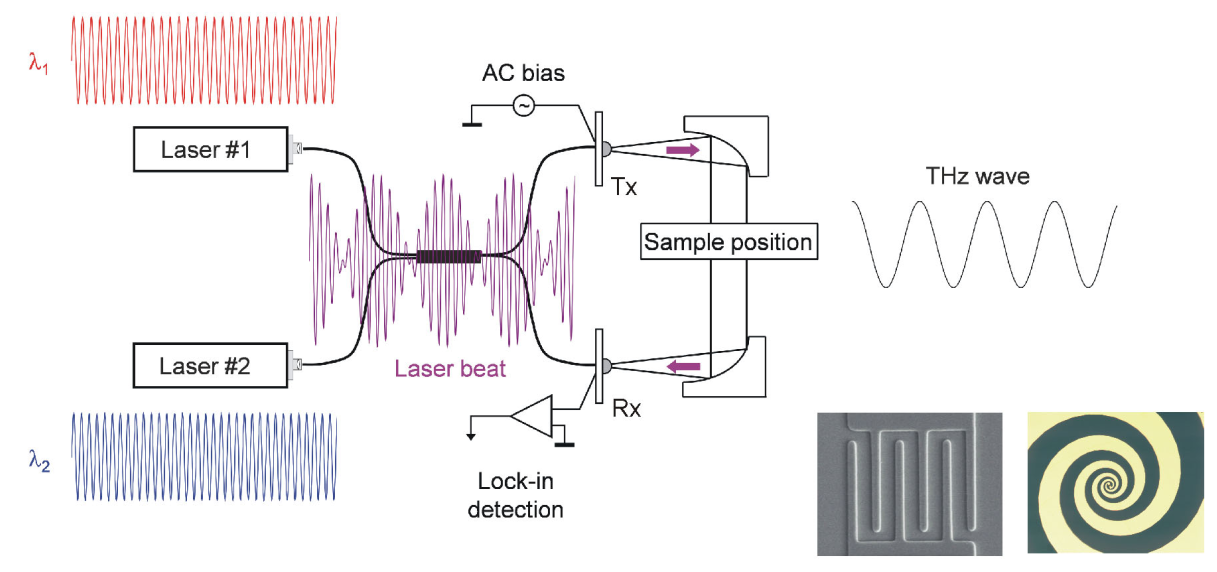}
    \includegraphics[width=0.25\linewidth]{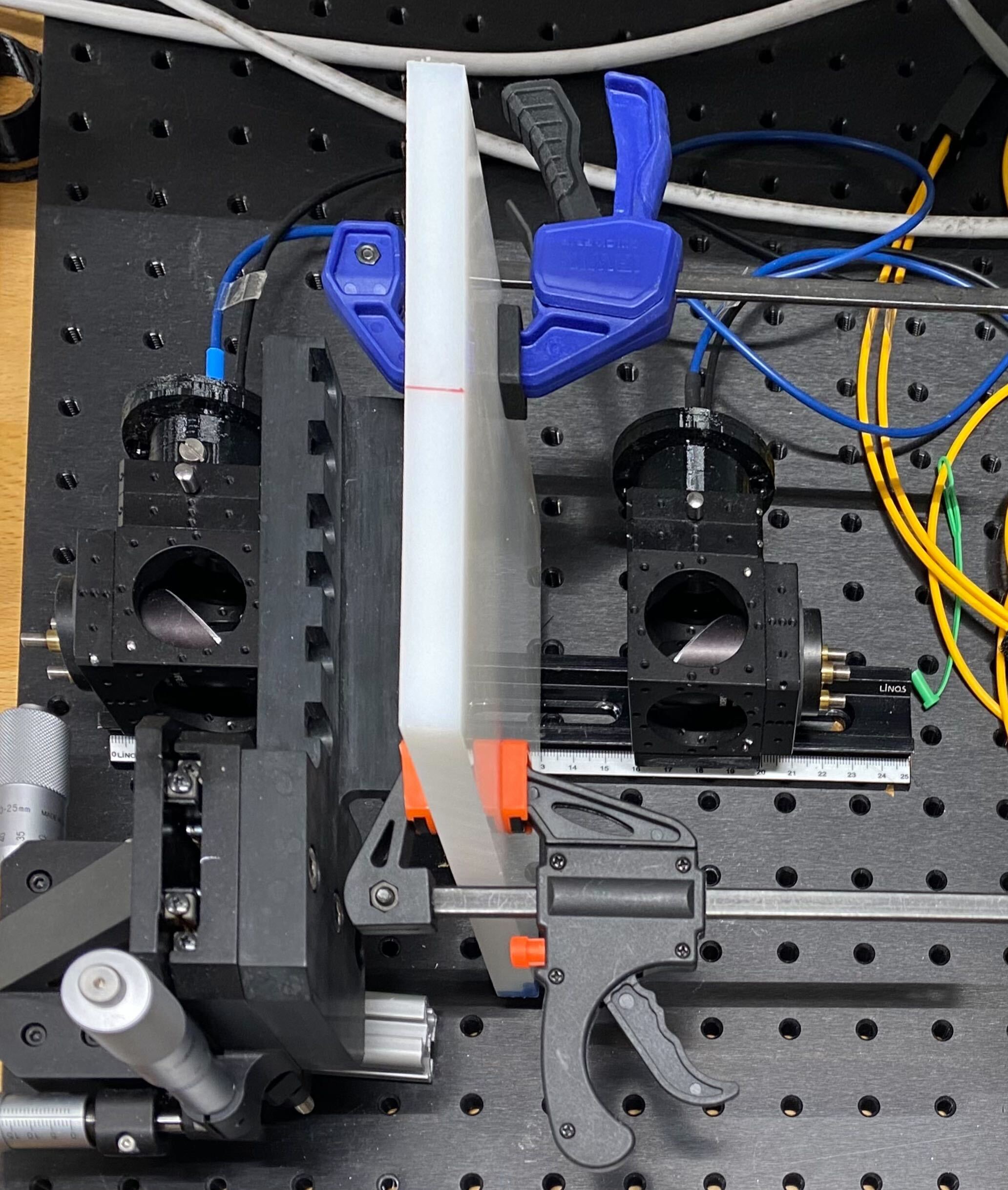}
    \caption{[Top] Diagram of continuous wave terahertz generation by optical heterodyning in photoconductors with lasers from \citenum{topt_man}. [Right] Photo of HDPE sample in TOPTICA transmission set up.}
    \label{fig:toptica_diagram}
\end{figure}

For sub-millimeter transmission measurements we use a TeraScan 1550 from TOPTICA. To generate the spectrum of sub-millimeter radiation, two near-infrared lasers are beat against a ``photomixer", which creates charge carriers in the semiconductor. Applying a bias voltage generates a current at the beat frequency between the two lasers, which is then coupled to free space with an antenna. A diagram of the terahertz wave production is provided in Figure \ref{fig:toptica_diagram}, with a photo of the HDPE sample in the measurement apparatus.

The raw data produced by the TeraScan are the measured photocurrent envelope in the photomixers. Dielectric properties of materials are determined in two stages. First, the frequency dependent electrical length is determined. The Fabry-Perot free spectral range ($n(\nu) \cdot d = c/(2\overline{\Delta \nu})$) is subtracted from the frequency shift of the interference maxima between the sample and free-space photocurrents ($(n-1)\cdot d$) to recover the thickness ($d$) and the index ($n(\nu)$) \cite{Roggenbuck2010,Roggenbuck2012}. The recovered thickness is then checked against the physical thickness measurement. Second, the extinction coefficient $k(\nu)$ is determined at each frequency by fitting the $T_\mathrm{max}$ envelope model described in Equation \ref{eq:tmax} to the measured envelope with a bounded root-finding algorithm. 
\begin{equation} \label{eq:tmax}
\begin{split}
    T_\mathrm{max}(k) &= \frac{(1-R)^2 e^{-\alpha d}}{(1-Re^{-\alpha d})^2} \\
    R &= \frac{(n-1)^2 +k^2}{(n=1)^2 +k^2}\\
    \alpha &= \frac{4\pi k \nu}{c}
\end{split}
\end{equation}
We integrate at each frequency in the TeraScan sweep for \SI{30}{\milli\second}, which gives us a dynamic range of approximately \SI{70}{\decibel}. A full sweep of the frequency range takes about 30 minutes to complete.

\section{Results}\label{sec:results}

\subsection{Scattering Scans} \label{sec:scans}

All scans reported here are a single high-angle scan of these materials. The extent of the scans range from $-15$ to $+79$ degrees in azimuth and $\pm$ 12 degrees in elevation, with a step size of 0.25 degrees. All scans are taken in a `left/right' raster pattern starting at azimuth $-15$ deg, elevation $-12$ deg and swept at constant elevation to azimuth 79 before stepping up in elevation and sweeping back. At each point we allow the robot to settle for 0.2 seconds before initiating a VNA frequency sweep. We report the magnitude of the measured scattered signal down to \SI{-80}{\decibel} to show the high dynamic range of our VNA. A full scan takes about nine hours to complete.

\subsubsection{Reference}
\begin{figure}[t]
    \centering
    \includegraphics[width=0.43\linewidth]{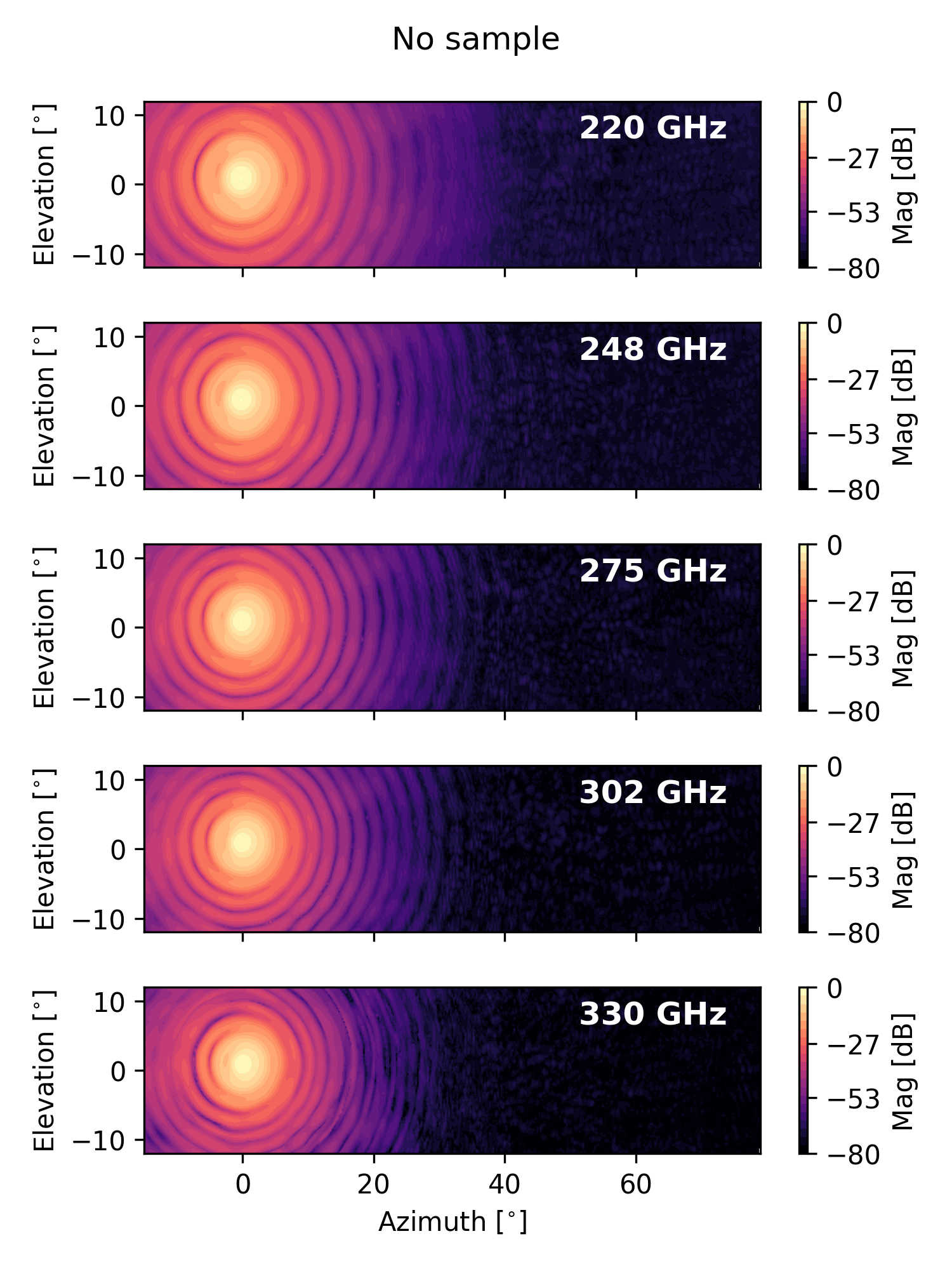}
    \includegraphics[width=0.43\linewidth]{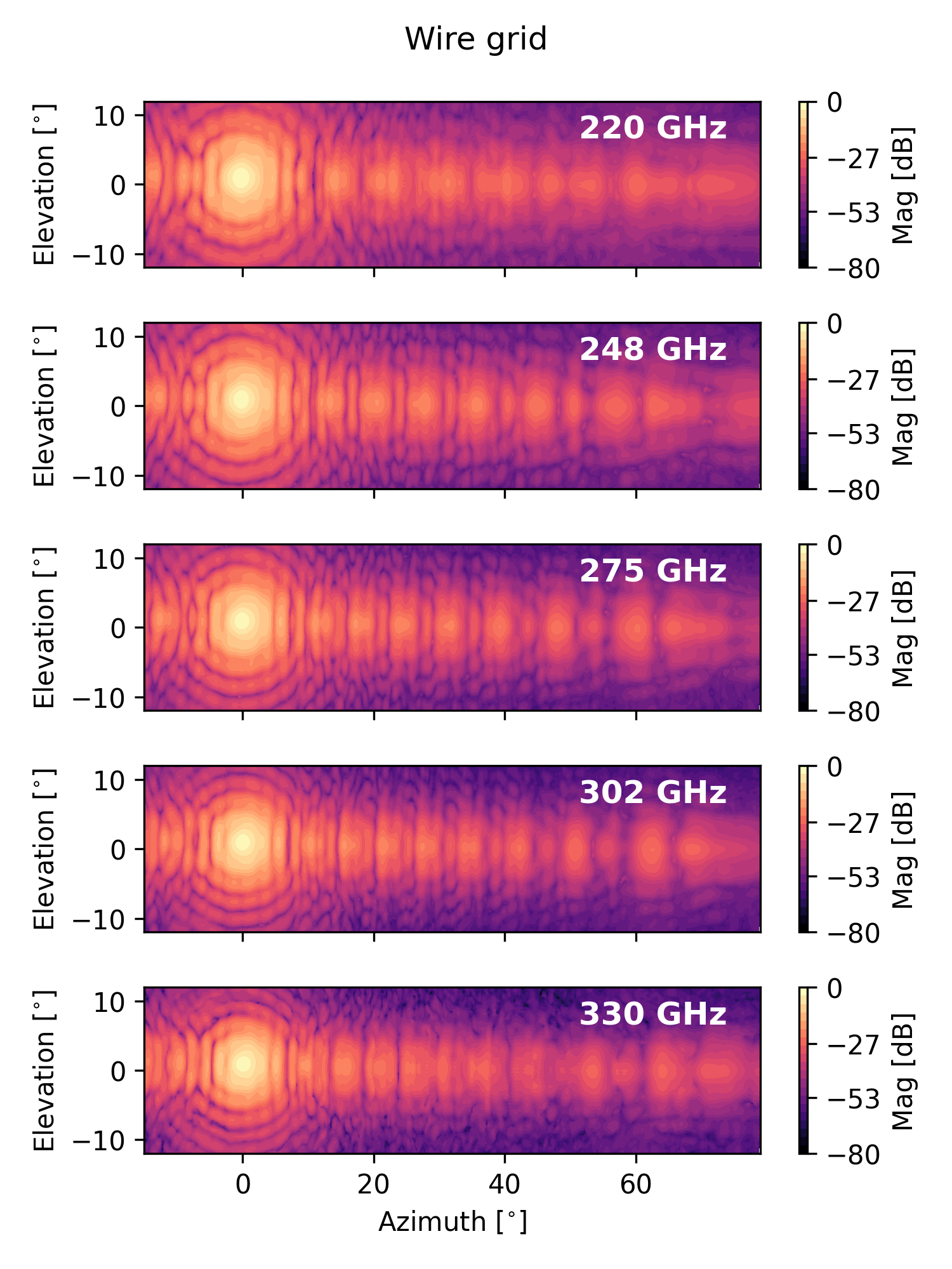}
    \caption{Reference scans for scatterometry with the J-band head (220-\SI{330}{\giga\hertz}). No sample [left] is the measurement apparatus with nothing in it, wire grid [right] is a sparse wire copper grid with \SIadj{1}{\centi\meter} spacing to produce maximal scattering.}
    \label{fig:ref}
\end{figure}

\begin{figure}[t]
    \centering
    \includegraphics[width=0.43\linewidth]{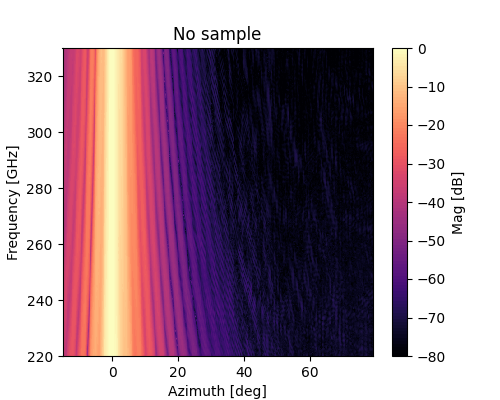}
    \includegraphics[width=0.43\linewidth]{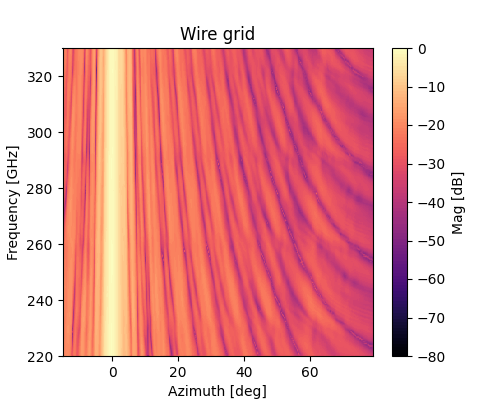}
    \caption{A slice at the center of the beam of reference scans over the full frequency range, normalized to the frequency at the beam center of the no sample measurement. [Left] No sample (empty) measurement. [Right] Wire grid measurement.}
    \label{fig:ref_f_v_a}
\end{figure}

The reference measurements of the set-up with no sample and with a sparse wire grid are shown in Figure \ref{fig:ref}; as expected, in the no sample case the signal quickly drops below the noise floor. The truncated beam also reduces in width at higher frequencies, as do the Airy rings. The \SIadj{330}{\giga\hertz} scan is potentially warped slightly by timegating issues with the band edge in all scans. The wire grid shows a high peak of diffracted power at the same elevation of the beam, which in turn speaks to how well aligned the wire grid was to the measurement surface. If the wire grid was tilted slightly with respect to the scan, we would expect the diffracted power to be tilted a similar degree in elevation.
 
We also show how scattered structure evolves with frequency by taking a slice through the scans at a constant elevation (shown in Figure \ref{fig:ref_f_v_a}). These plots are normalized in frequency to the beam center measurement of the no sample case (see Figure  \ref{fig:processing}). Again, each reference measurement evolves as expected with frequency.

\begin{figure}[th!]
    \centering
    \includegraphics[width=0.45\linewidth]{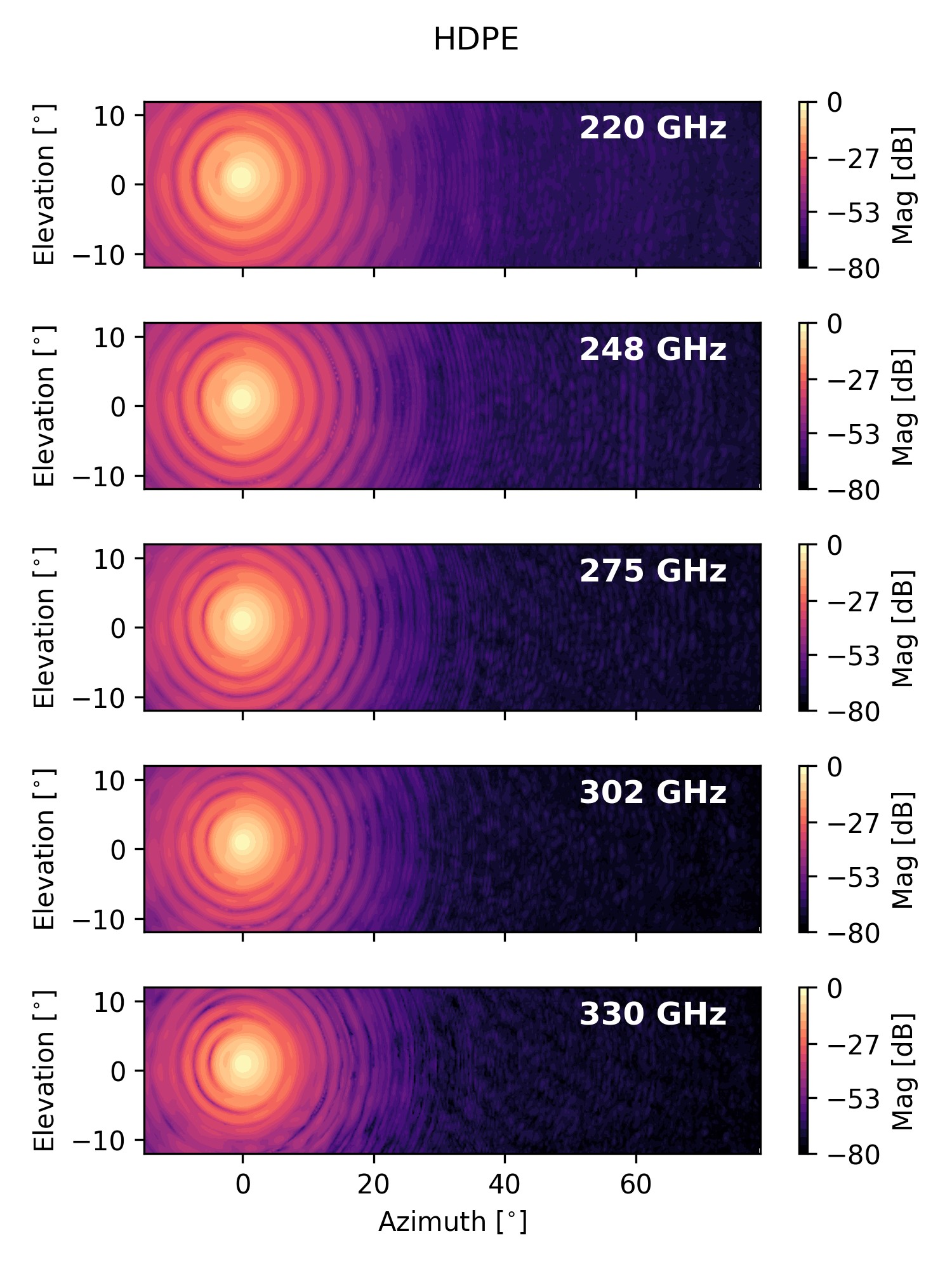}
    \includegraphics[width=0.45\linewidth]{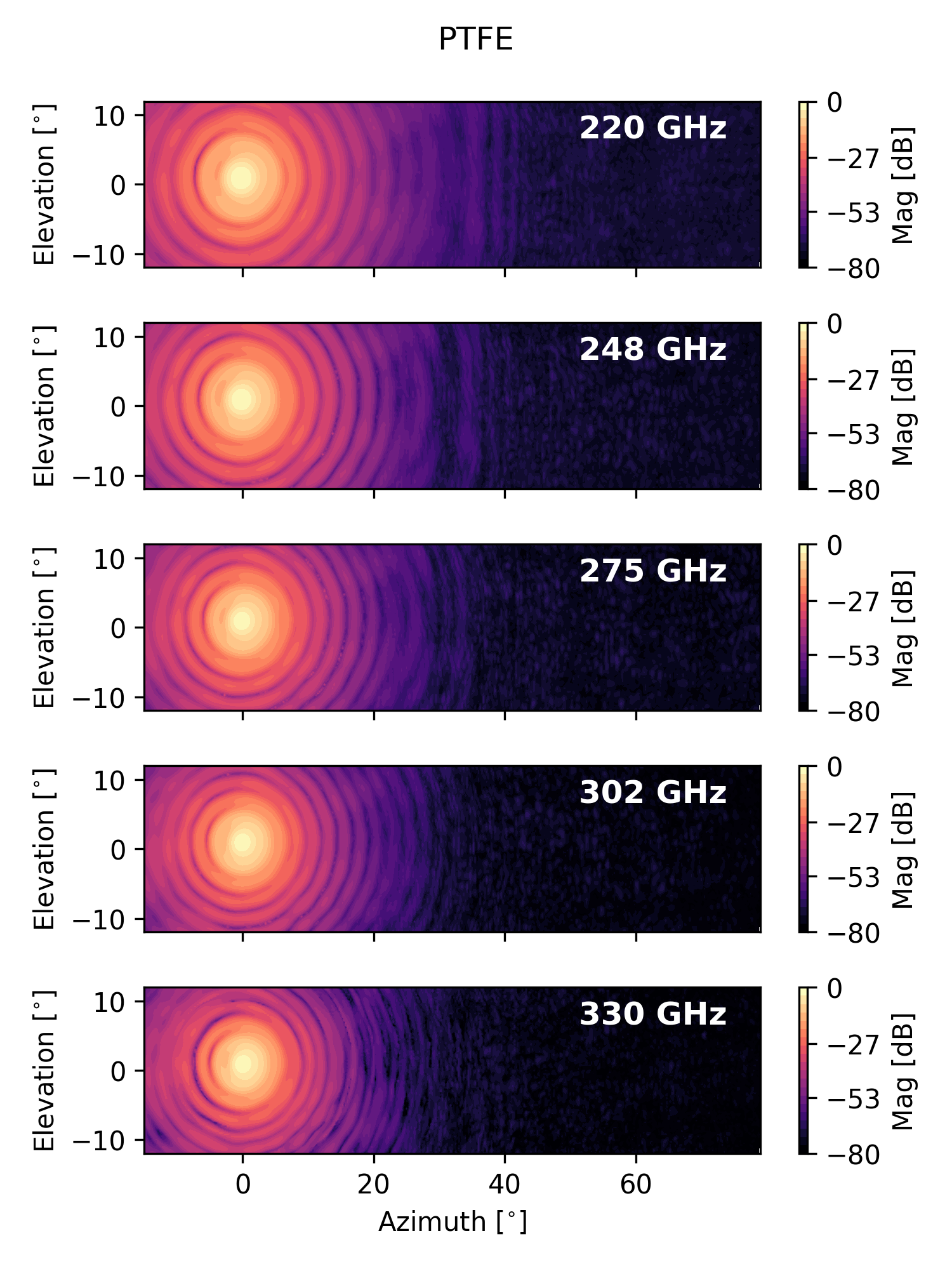}
    \includegraphics[width=0.45\linewidth]{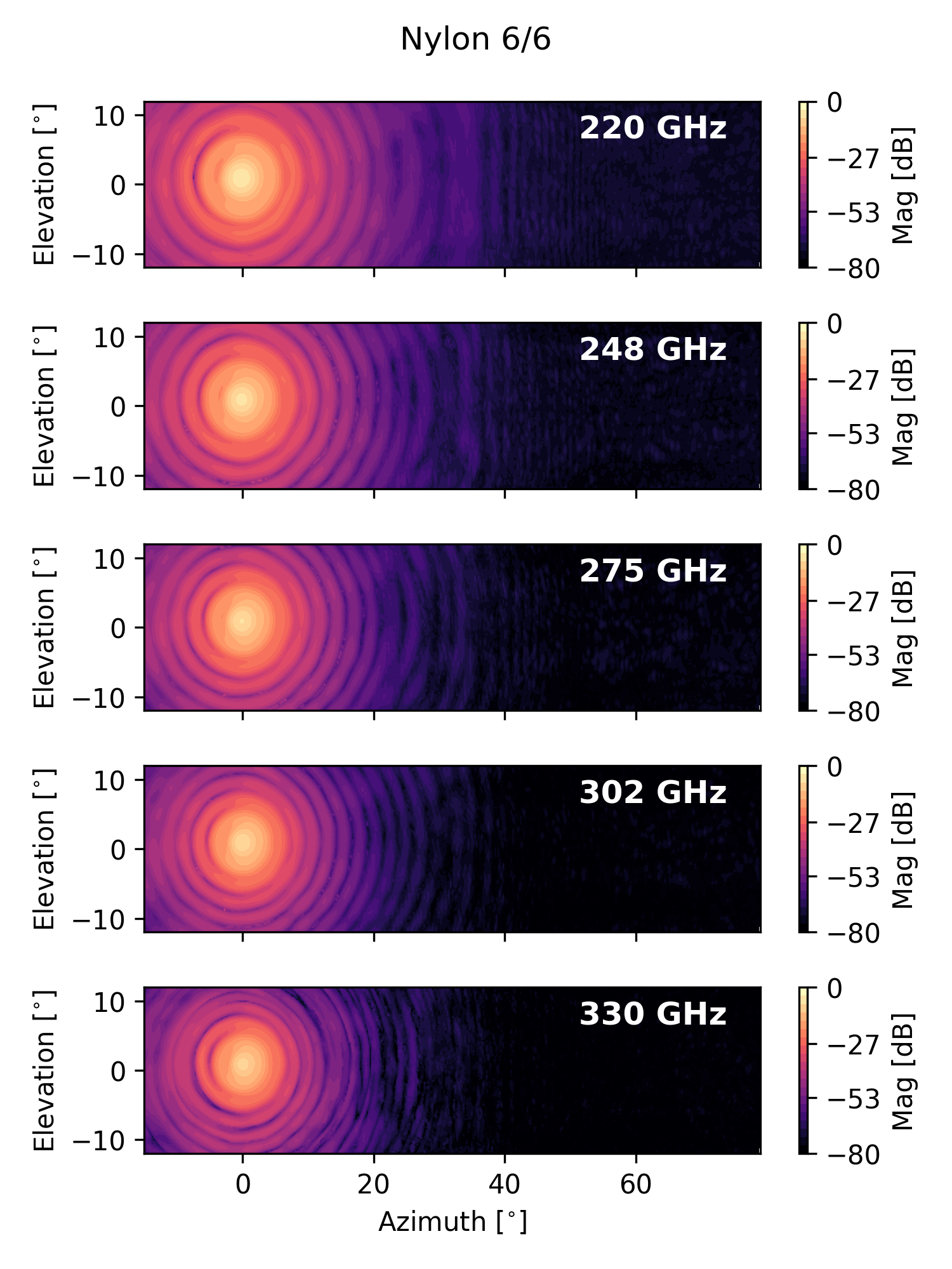}
    \caption{Bulk polymer scatterometry scans with the J-band head (220-\SI{330}{\giga\hertz}). Bulk HDPE (1/2", \SIadj{12}{\milli\meter} thick) [top left] is used primarily for lenses and windows, while bulk PTFE (1/4", \SIadj{6}{\milli\meter} thick) [top right] and nylon (1/4", \SIadj{6}{\milli\meter} thick) [bottom] have been used for IR filters.}
    \label{fig:bulk}
\end{figure}

\clearpage

\subsubsection{Bulk polymers}

\begin{figure}[t!]
    \centering
    \includegraphics[width=0.45\linewidth]{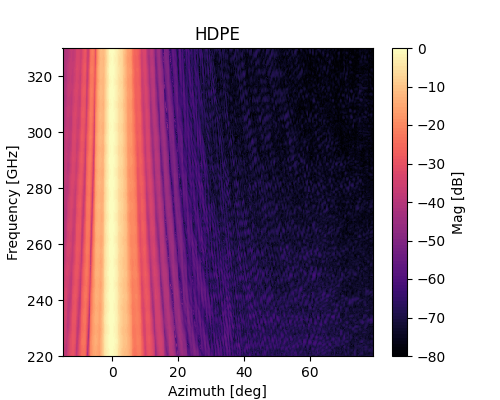}
    \includegraphics[width=0.45\linewidth]{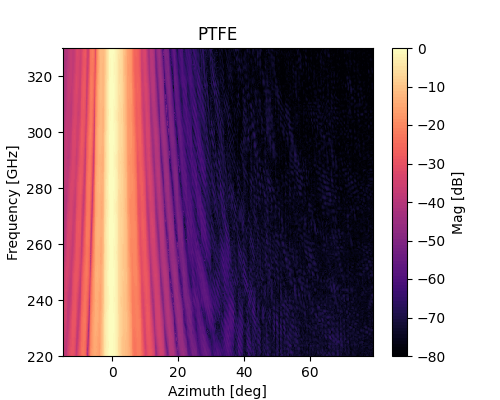}
    \includegraphics[width=0.45\linewidth]{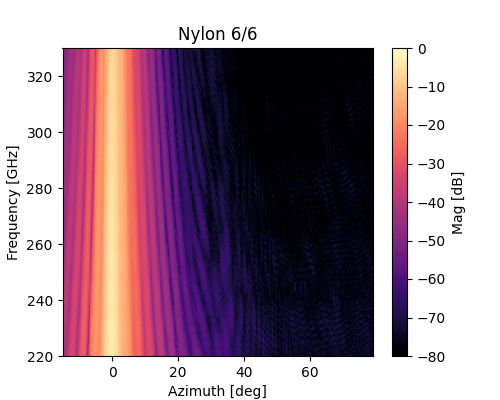}
    \caption{A slice at the center of the beam of bulk polymer scans over the full frequency range, normalized to the frequency at the beam center of the no sample measurement. [Top left] HDPE (1/2", \SIadj{12}{\milli\meter} thick), [top right] PTFE (1/4", \SIadj{6}{\milli\meter} thick), [bottom] nylon 6/6 (1/4", \SIadj{6}{\milli\meter} thick).}
    \label{fig:bulk_f_v_a}
\end{figure}

We see essentially no scattered power from the bulk polymers shown in Figure \ref{fig:bulk}. There is a sharp feature just below 40 degrees that is common between all samples; this is likely diffraction off the sample edge, as the shape is very flat and diminishes at high frequency as the beam width decreases.

The constant elevation slices for these polymers shown in Figure \ref{fig:bulk_f_v_a} also show this common sharp feature disappearing at higher frequencies. Other features associated with the optical properties of the materials are visible, such as the frequency fringing in the main beam in the HDPE (associated with Fabre-Perot resonance within the sample) and how nylon reduces power significantly in the main beam at higher frequencies.

HDPE does show a slight increase in signal at high angles compared to the other bulk polymers. This may be because the HDPE sample was twice as thick as the other samples. Otherwise, we see essentially no scattered power through these bulk polymers.

\subsubsection{Structured surface}

\begin{figure}[t]
    \centering
    \includegraphics[width=0.45\linewidth]{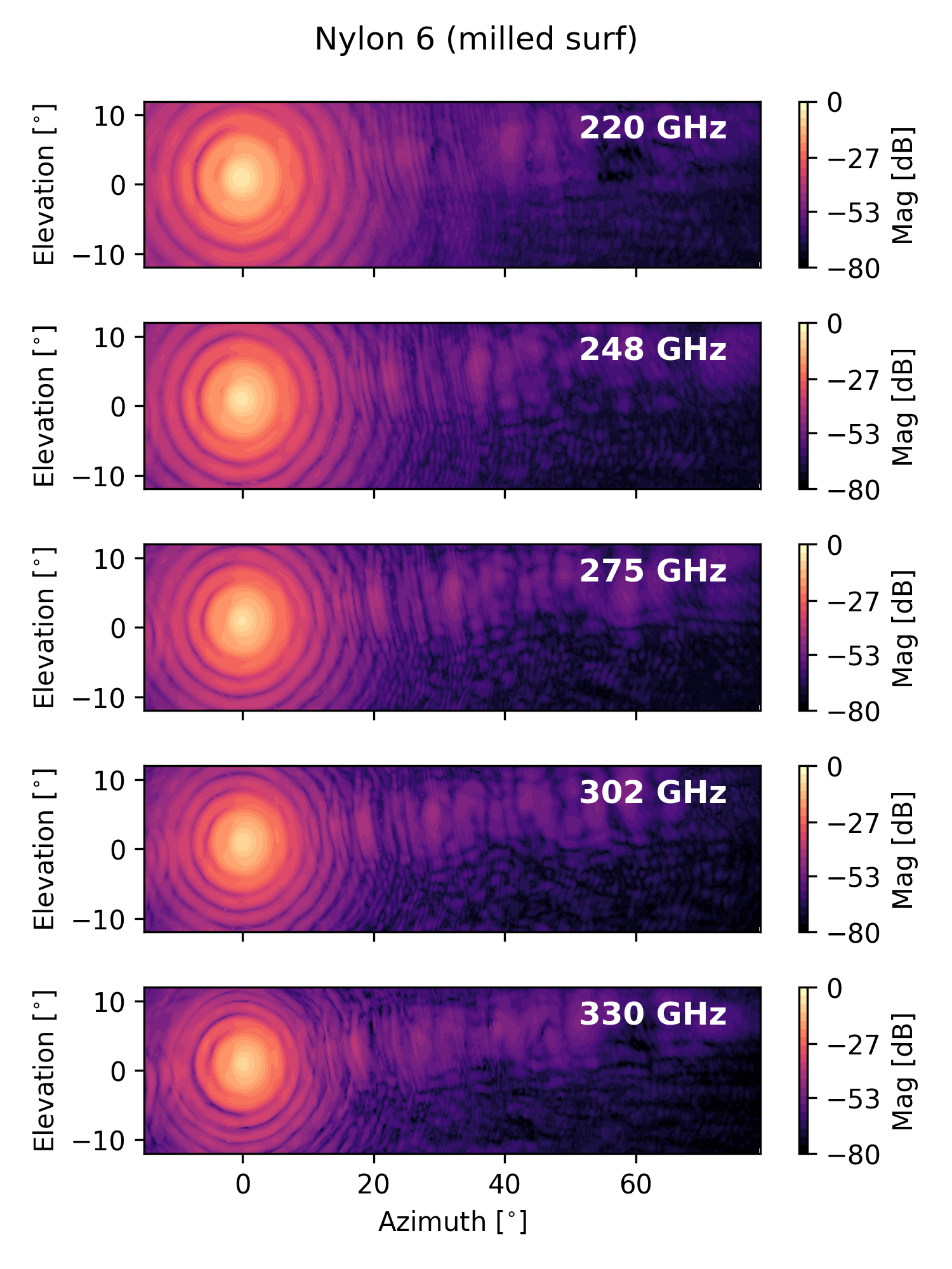}
    \includegraphics[width=0.5\linewidth]{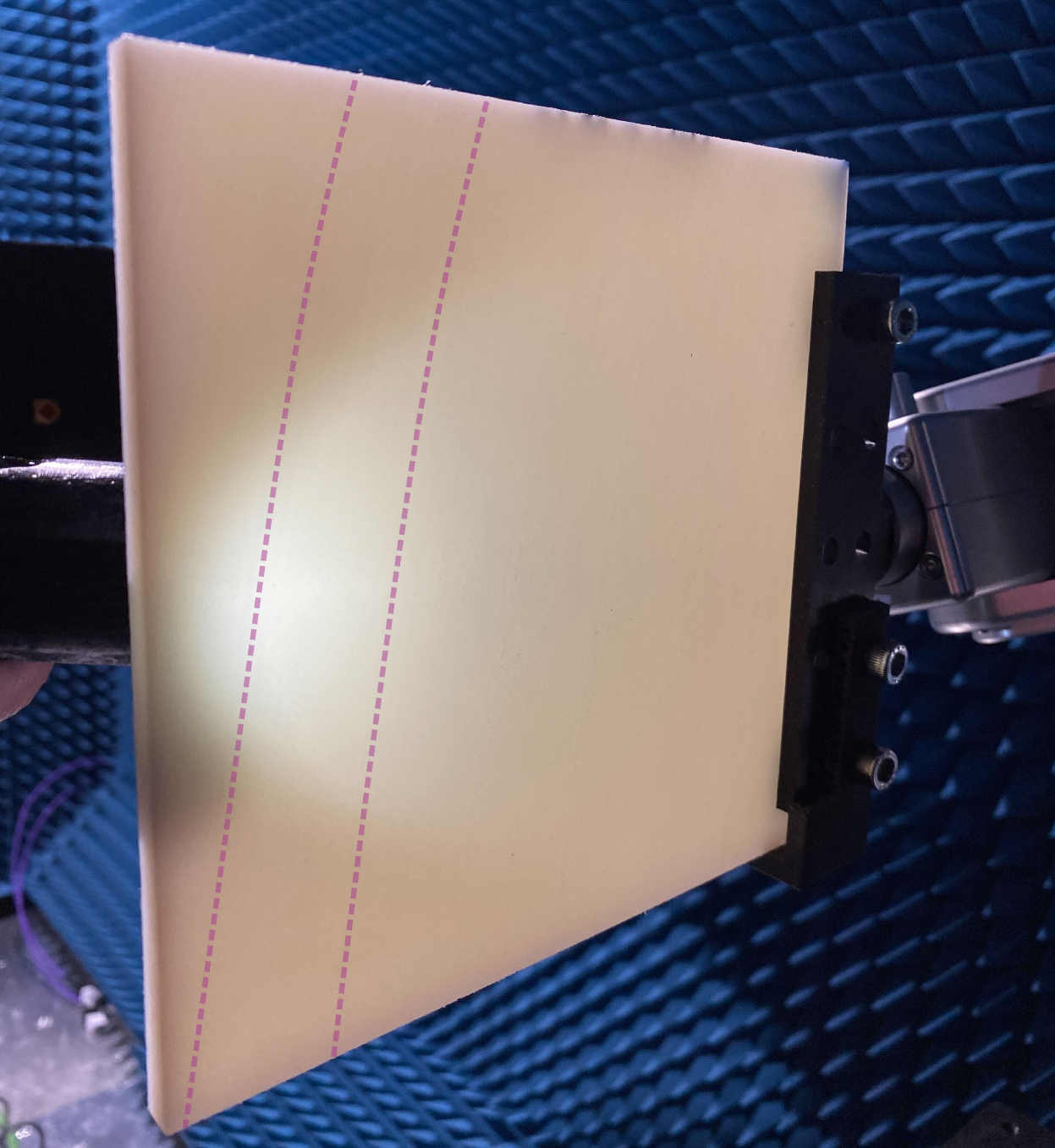}
    \caption{[Left] Nylon 6 scatterometry scan with the J-band head (220-\SI{330}{\giga\hertz}). [Right] Photo of the nylon 6 sample in the scatterometry set up. Purple dashed lines have been superimposed to align with the milled surface structure present on the sample.}
    \label{fig:nylon6}
\end{figure}

\begin{figure}[t]
    \centering
    \includegraphics[width=0.5\linewidth]{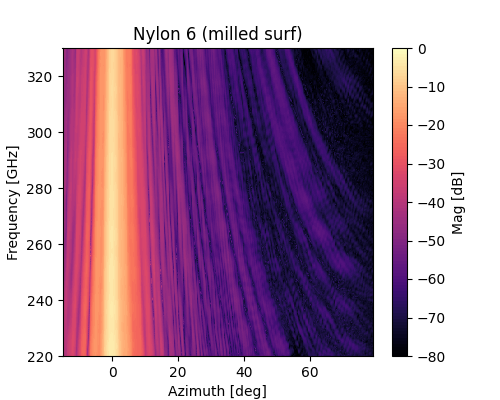}
    \caption{A slice at the center of the beam of the nylon 6 scan over the full frequency range, normalized to the frequency at the beam center of the no sample measurement.}
    \label{fig:nylon6_f_v_a}
\end{figure}

All of our nylon 6 samples arrived with clear fly-cut milled surface features. These grooves are all parallel with a slight curve and were angled slightly compared to the measurement plane, as shown with the dashed lines superimposed on the photo in Figure \ref{fig:nylon6}. The peak height of the grooves are of order 10 microns.

These grooves clearly caused surface scattering of the beam perpendicular to the groove direction, as shown in Figures \ref{fig:nylon6} and \ref{fig:nylon6_f_v_a}. The diffracted power show similar frequency dependence to the wire grid measurements, as one may expect scattering sourced from parallel line surface structure to behave.

\subsubsection{Polymer foam stacks}

\begin{figure}[t]
    \centering
    \includegraphics[width=0.45\linewidth]{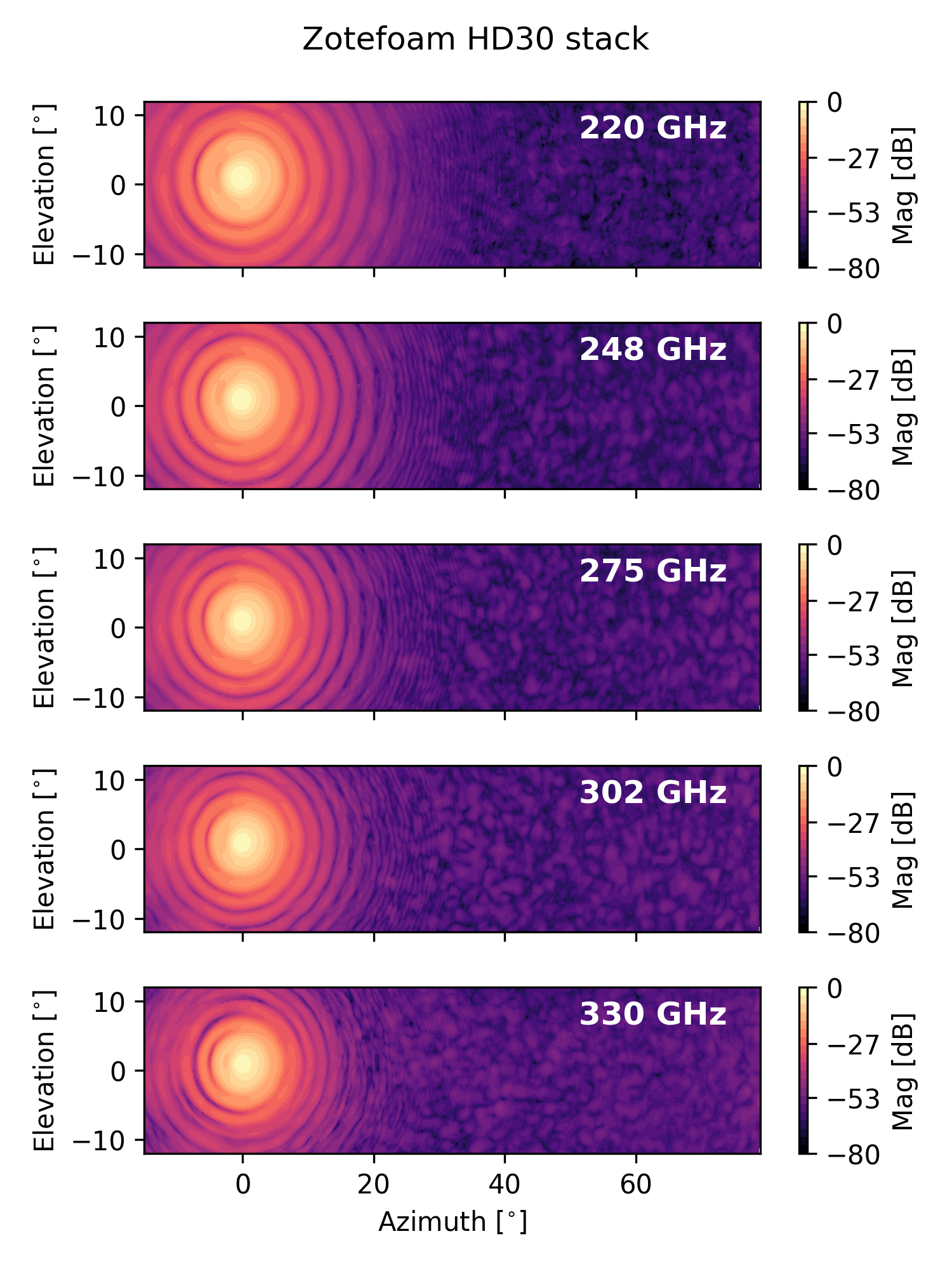}
    \includegraphics[width=0.45\linewidth]{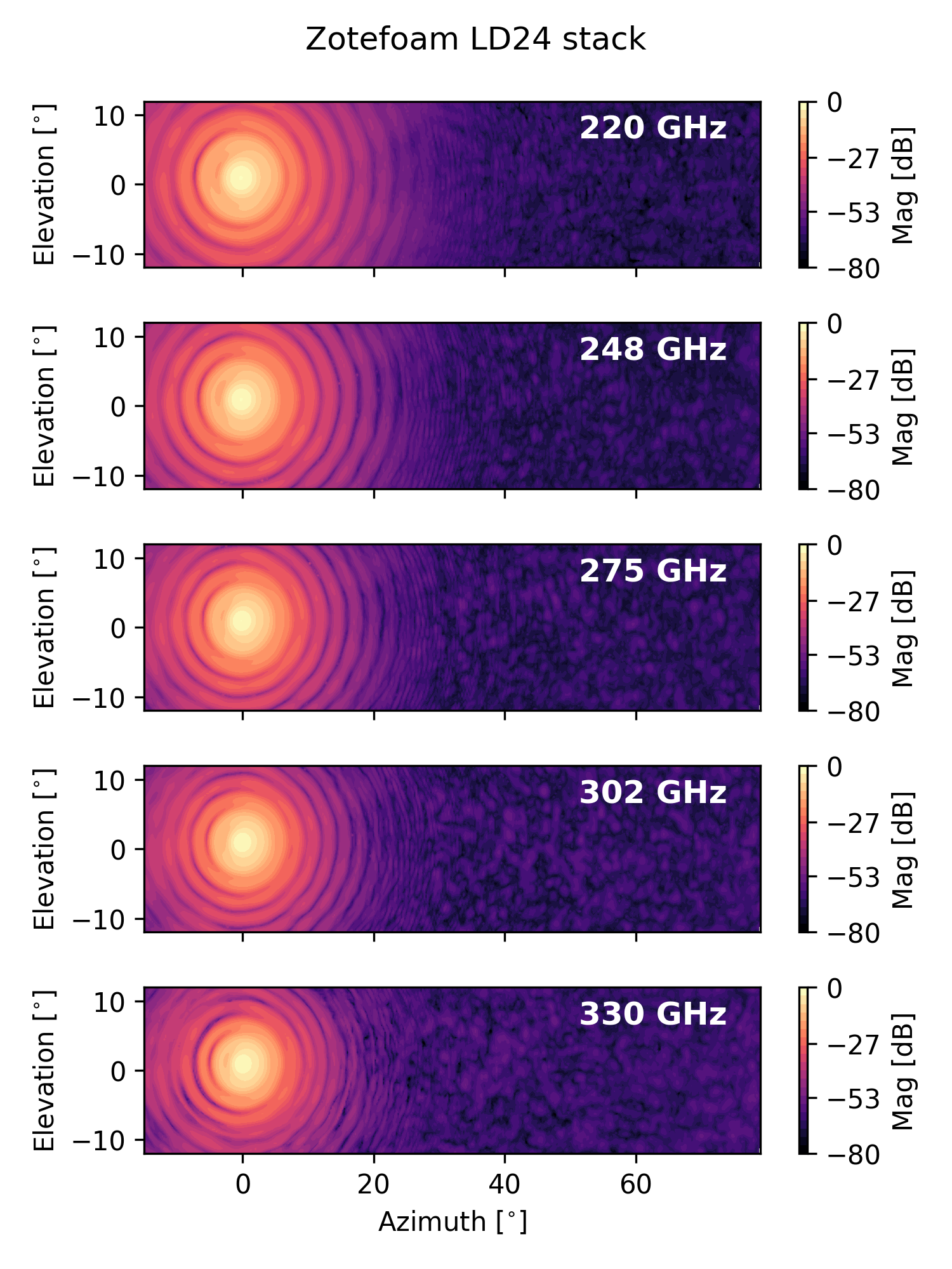}
    \includegraphics[width=0.6\linewidth]{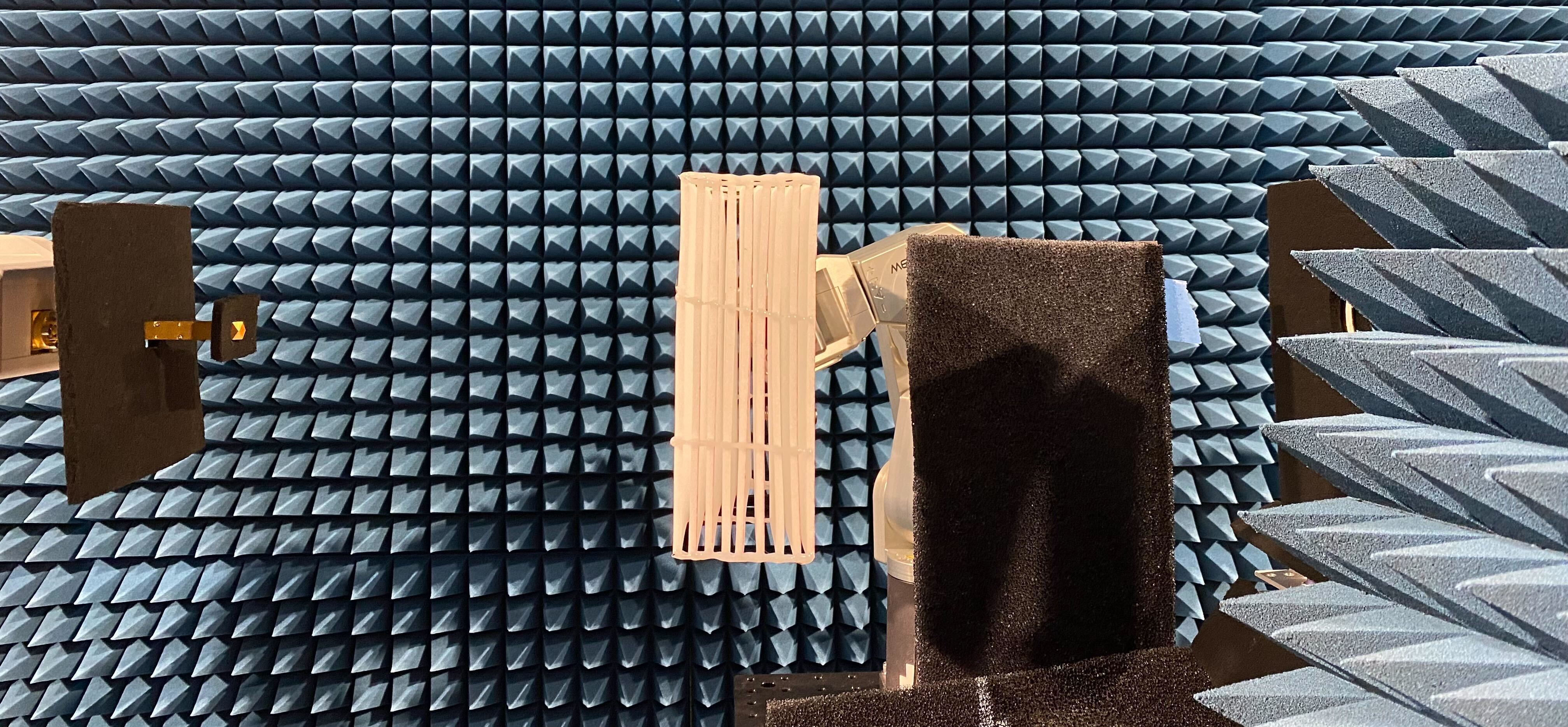}
    \caption{Zotefoam stack scans with the J-band head (220-\SI{330}{\giga\hertz}). The scan of the stack of nine HD30 \SI{3}{\milli\meter} sheets spaced approximately \SI{3}{\milli\meter} apart [top left], the scan of a stack of nine LD24 \SIadj{3}{\milli\meter} thick sheets spaced approximately \SI{3}{\milli\meter} apart [top left], and a photo of the stack in the sample holder [bottom].}
    \label{fig:zotefoam}
\end{figure}

\begin{figure}
    \centering
    \includegraphics[width=0.45\linewidth]{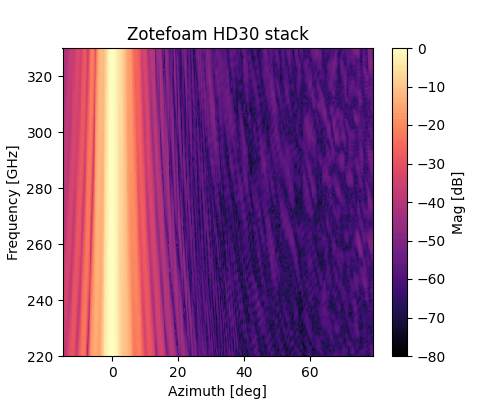}
    \includegraphics[width=0.45\linewidth]{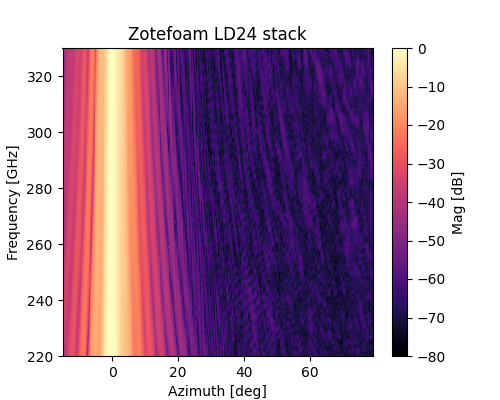}
    \caption{A slice at the center of the beam of Zotefoam scans over the full frequency range, normalized to the frequency at the beam center of the no sample measurement.}
    \label{fig:zote_f_v_a}
\end{figure}

The Zotefoam stacks were constructed by cutting nine 16 by \SI{16}{\centi\meter} squares from the same sheet. The smaller squares were then stacked with \SIadj{3}{\milli\meter} thick cardboard spacers. The edges were glued with ribbons of hot glue to add structure and keep the space between layers before the spacers were removed. This spacing was modeled after the RT-MLI filter described in Goldfinger, et al., 2022 \cite{Goldfinger2022}.

These filter stacks show a consistent excess of scattered power across the measured frequency range, as seen in Figure \ref{fig:zotefoam}. The excess is relatively homogeneous across elevation, though there are distinct `bubble' structures that can be followed inward at higher frequencies, as seen in the elevation slices in Figure \ref{fig:zote_f_v_a}. This may be expected given the relatively large cell size within both HD30 and LD24 of order \SI{400}{\micro\meter} \cite{Alex2026foam}. The LD24 scatters slightly less power than the HD30 stack. This may be due to cell size differences between the two foams, sheet variation, or potentially LDPE's slightly lower index of refraction compared to HDPE.

The type of scattering may explain the homogeneous scattered power across the scan. The relative particle cross section can be parameterized by the ratio $x = \frac{\pi d}{\lambda}$
where $x$ is a dimensionless parameter, $d$ is the diameter of the particle and $\lambda$ is the wavelength of the incident light. At these frequencies and with a particle diameter of \SI{400}{\micro\meter}, $x$ is between 0.92 and 1.38, which is expected to be in the Mie scattering regime. Mie scattering has a relatively homogeneous scattering kernel at these angles. Variations across the sheet in cell size and integrated internal structure may also produce Rayleigh scattering, though we expect to a lesser extent. 

The relatively high level of homogeneous scattered power out to high angles from these foams is of particular concern given their current use as RT-MLI, which is typically very high in the optical chain. Therefore, a majority of this power is expected to terminate on the warm forebaffles. We will explore the potential impact of this in future work.

\subsubsection{Integrated Power}

\begin{figure}
    \centering
    \includegraphics[width=0.57\linewidth]{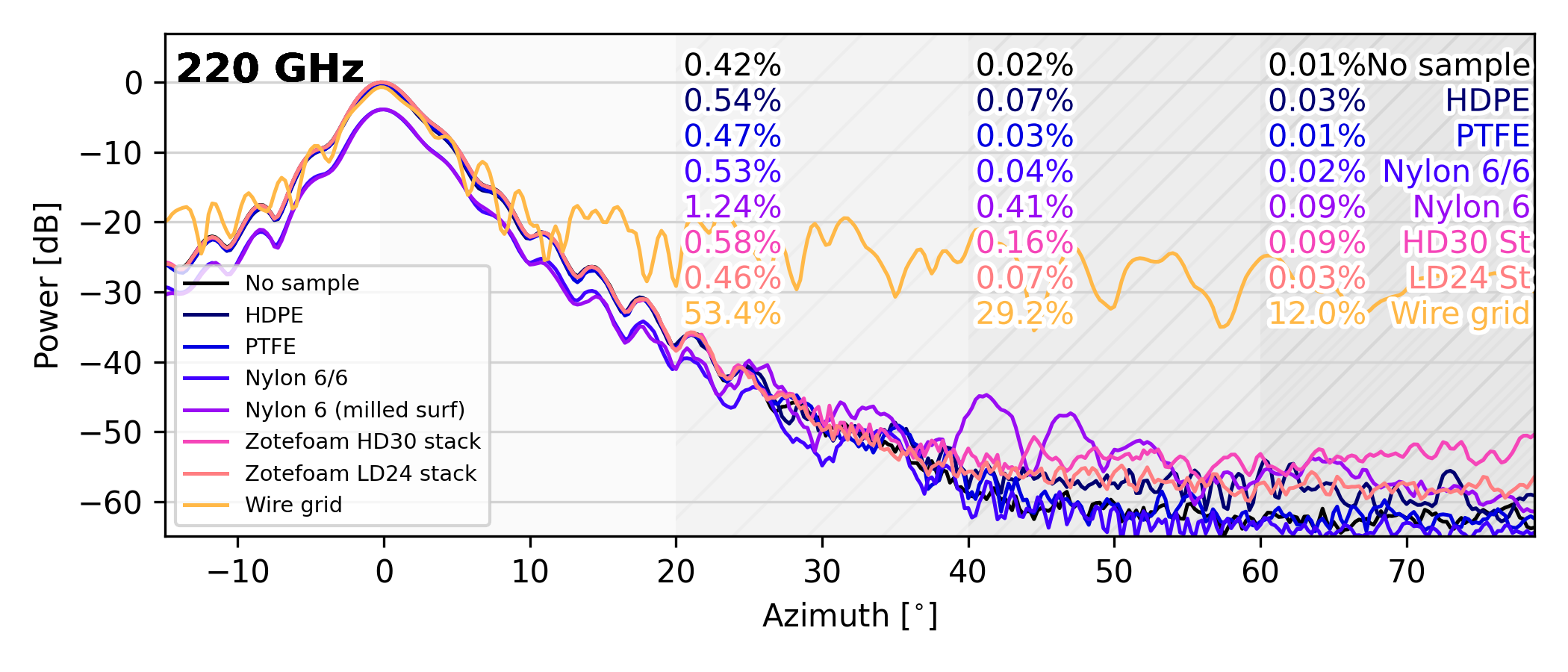}
    \includegraphics[width=0.57\linewidth]{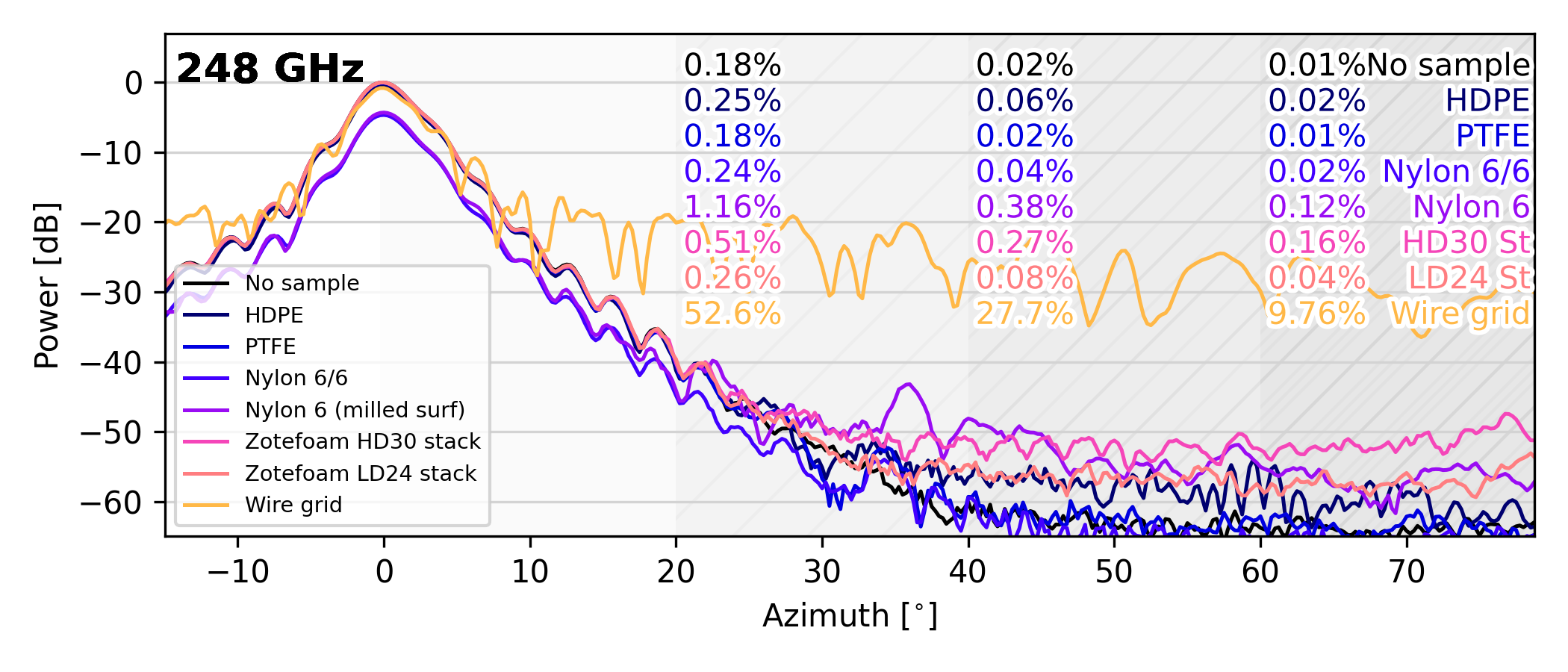}
    \includegraphics[width=0.57\linewidth]{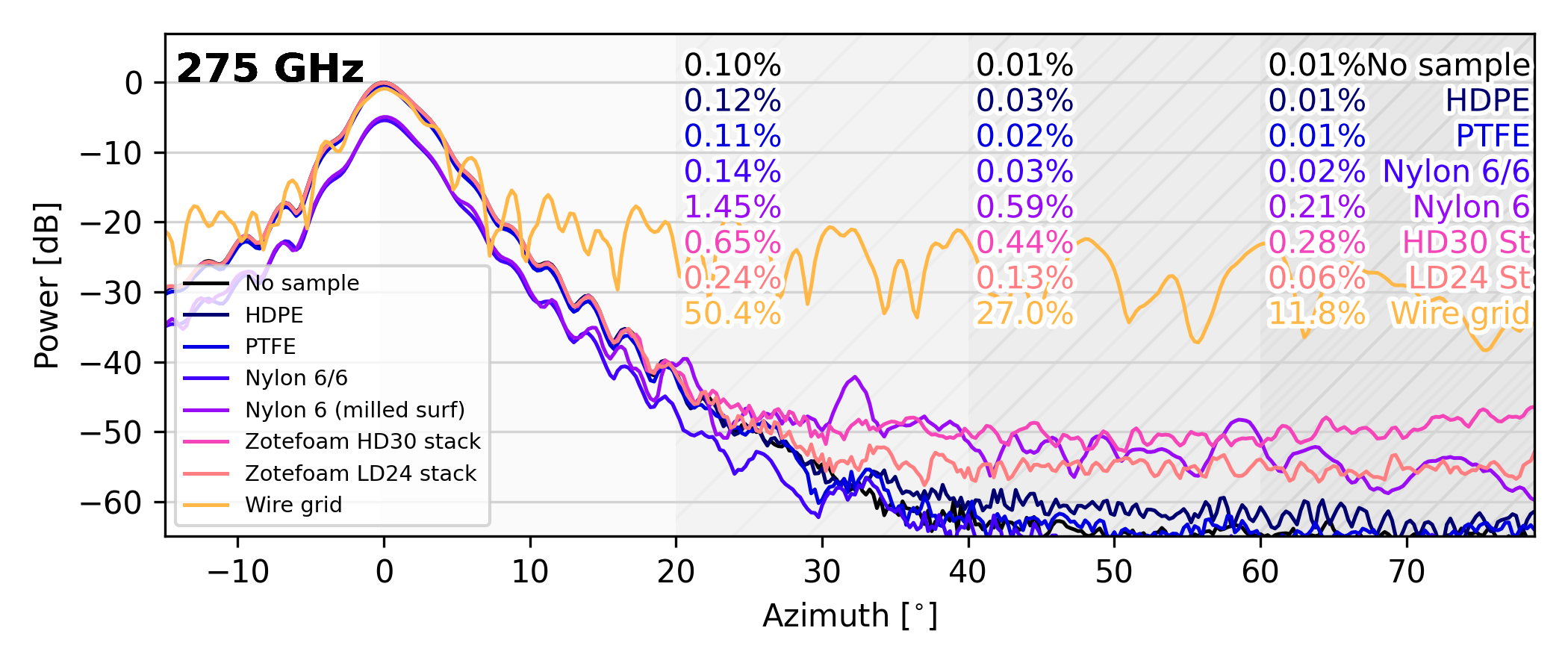}
    \includegraphics[width=0.57\linewidth]{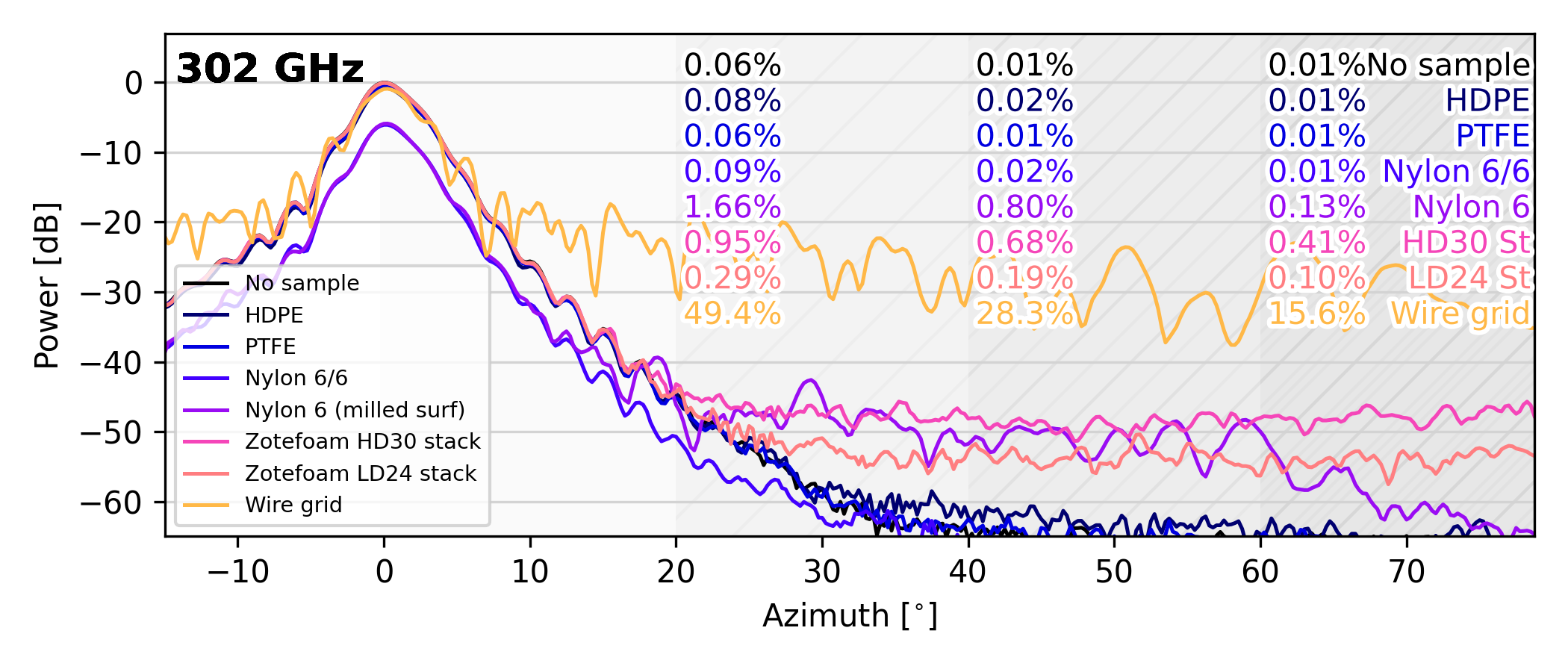}
    \includegraphics[width=0.57\linewidth]{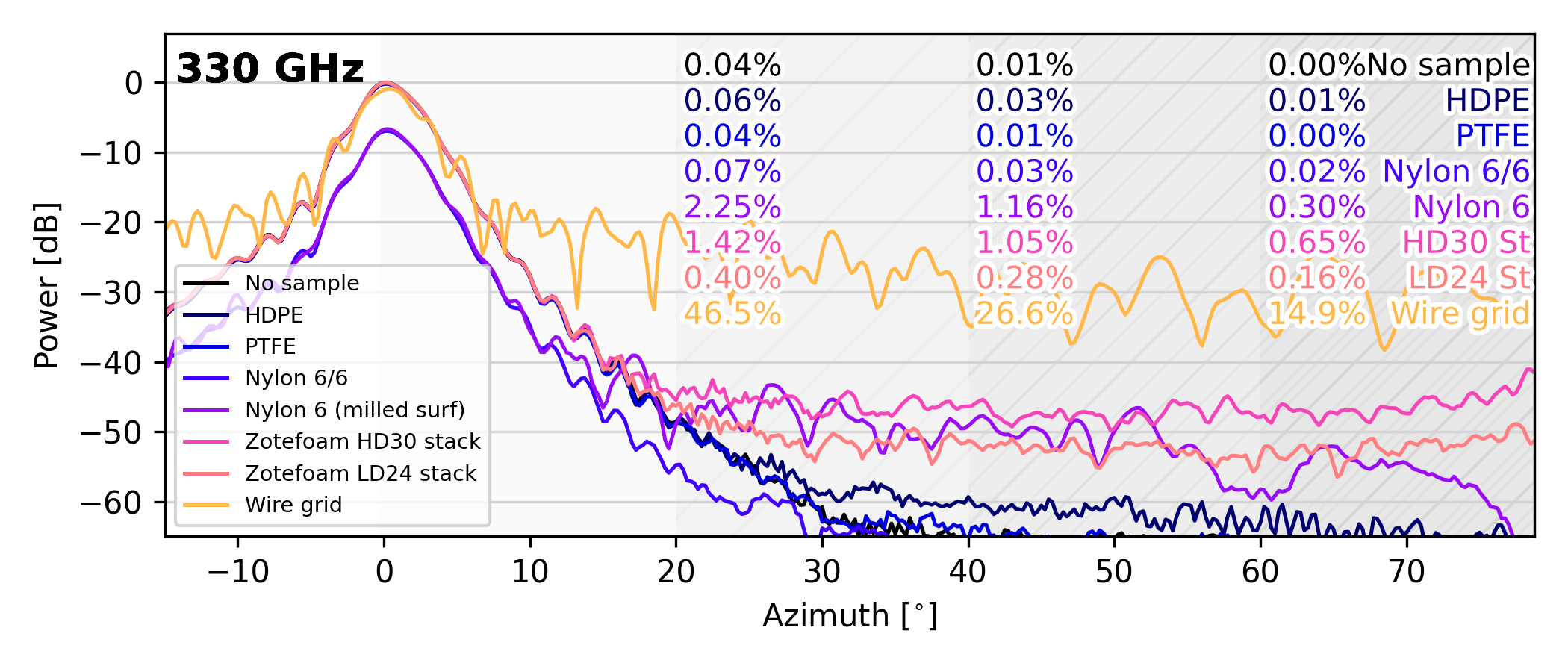}
    \caption{Scattered power in azimuth integrated over elevation and normalized to the peak height of the no sample (empty) measurement at the five frequencies shown in Figures \ref{fig:ref}, \ref{fig:bulk}, \ref{fig:nylon6}, and \ref{fig:zotefoam}.}
    \label{fig:int_pow}
\end{figure}

To directly compare the scattered power between materials we integrate the scans over elevation ($\phi$) at each point in azimuth ($\text{az}$) by
\begin{equation}
    P = \int_{-12^{\circ}}^{12^{\circ}} P(\text{az},\text{el}) ~d\text{el}
\end{equation}
and then normalizing to the integrated peak height of the no sample measurement. The integrated power over elevation curves are shown in Figure \ref{fig:int_pow}.

We estimate the fraction of power scattered out to higher angles by a material like:
\begin{equation}
    \begin{split}
        P_{} = \big(\frac{\int_{\text{az}'}^{79^{\circ}}P(\text{az}) \sin(\text{az}) ~d\text{az}}{\int_0^{79^{\circ}}P(\text{az}) \sin(\text{az}) ~d\text{az}}\big)100\%
    \end{split}
\end{equation}
where $\text{az}'$ is the angle past which we are integrating power (in Figure \ref{fig:int_pow} $\theta'$ is 20$^{\circ}$, 40$^{\circ}$, and 60$^{\circ}$), $\text{az}$ is azimuth angle, and $P(\text{az})$ is the power integrated over elevation at that that azimuth. This allows us to compare roughly how much of the main beam is getting scattered out to high angles, though the integration assumes a certain amount of homogeneity to the scattered power, which may not be the case for certain samples such as the wire grid and the nylon 6 sample with a structured surface.

The materials that push a significant amount of power out to high angles are the nylon 6 with the milled surface structure, the Zotefoam stacks and the wire grid, as anticipated from the scans. Also apparent in these plots is the reduction of power in the main beam from absorption in the nylon samples, and how that absorption gets worse at higher frequencies. We will discuss this further in the next section.

\subsection{Transmission}

\begin{figure}
    \centering
    \includegraphics[width=0.7\linewidth]{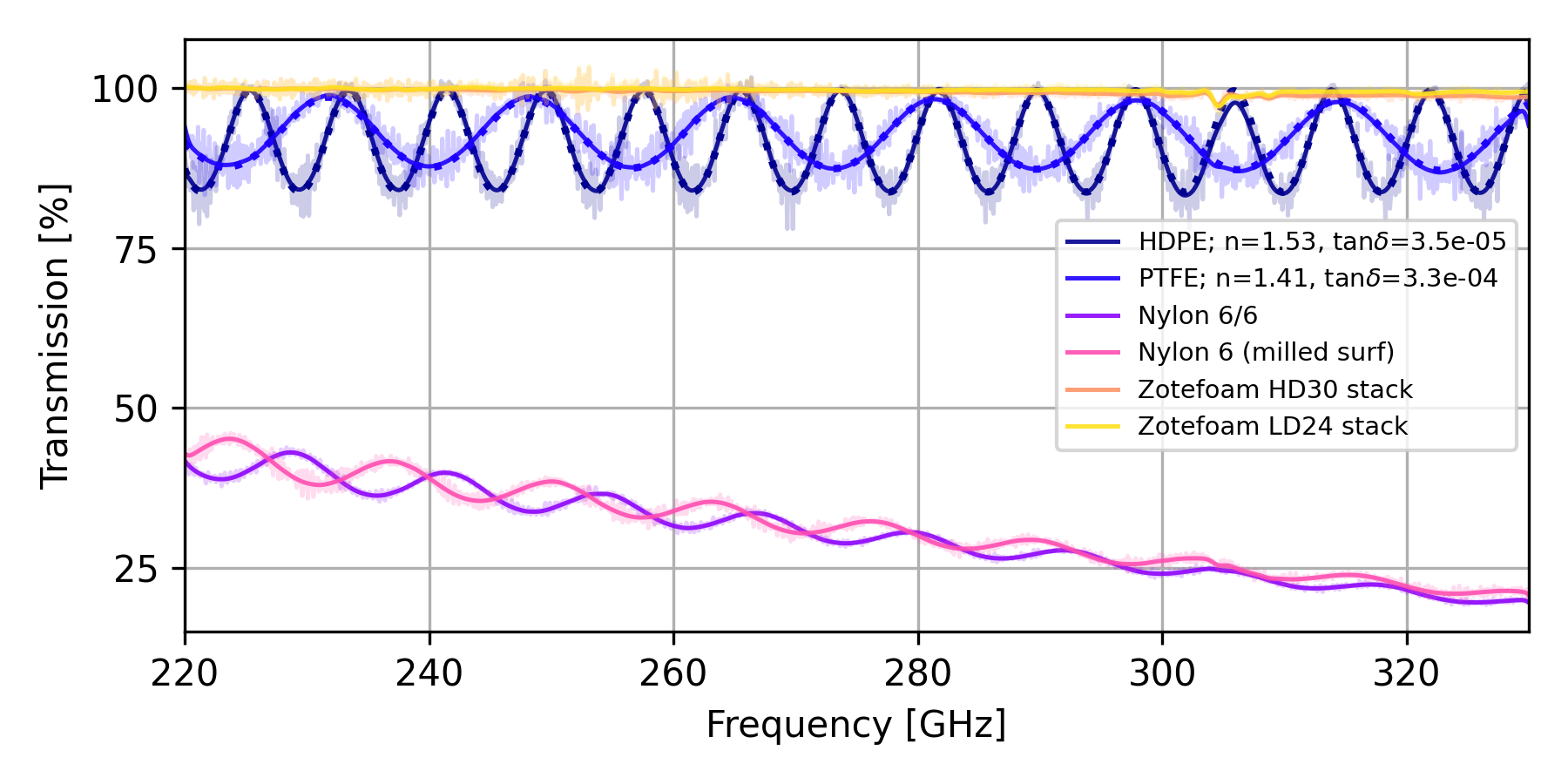}
    \caption{Transmission over frequency at the beam center of the scatterometry scans for HDPE, PTFE, nylon 6/6, nylon 6 and both Zotefoam stacks. Lighter alpha lines are raw VNA spectra, solid lines are time-gated spectra and dotted lines are fits to the transmission.}
    \label{fig:vna_trans}
\end{figure}

We show the VNA frequency spectra for the material measurement from the peak of their scans in Figure \ref{fig:vna_trans}. The data are normalized by the no sample measurement as described in Section \ref{sec:vna}: raw data are shown with lighter alpha, while timegated data are the darker lines. For the two low loss materials (HDPE and PTFE) we also fit a simple transmission model (via the transfer matrix method) to the timegated data to recover the complex index of refraction, the parameters of which are reported in the legend. This model does not currently account for a variable complex index across the measured frequency range, so it was not possible to use it to fit to the variably lossy nylon data.

\begin{figure}
    \centering
    \includegraphics[width=0.8\linewidth]{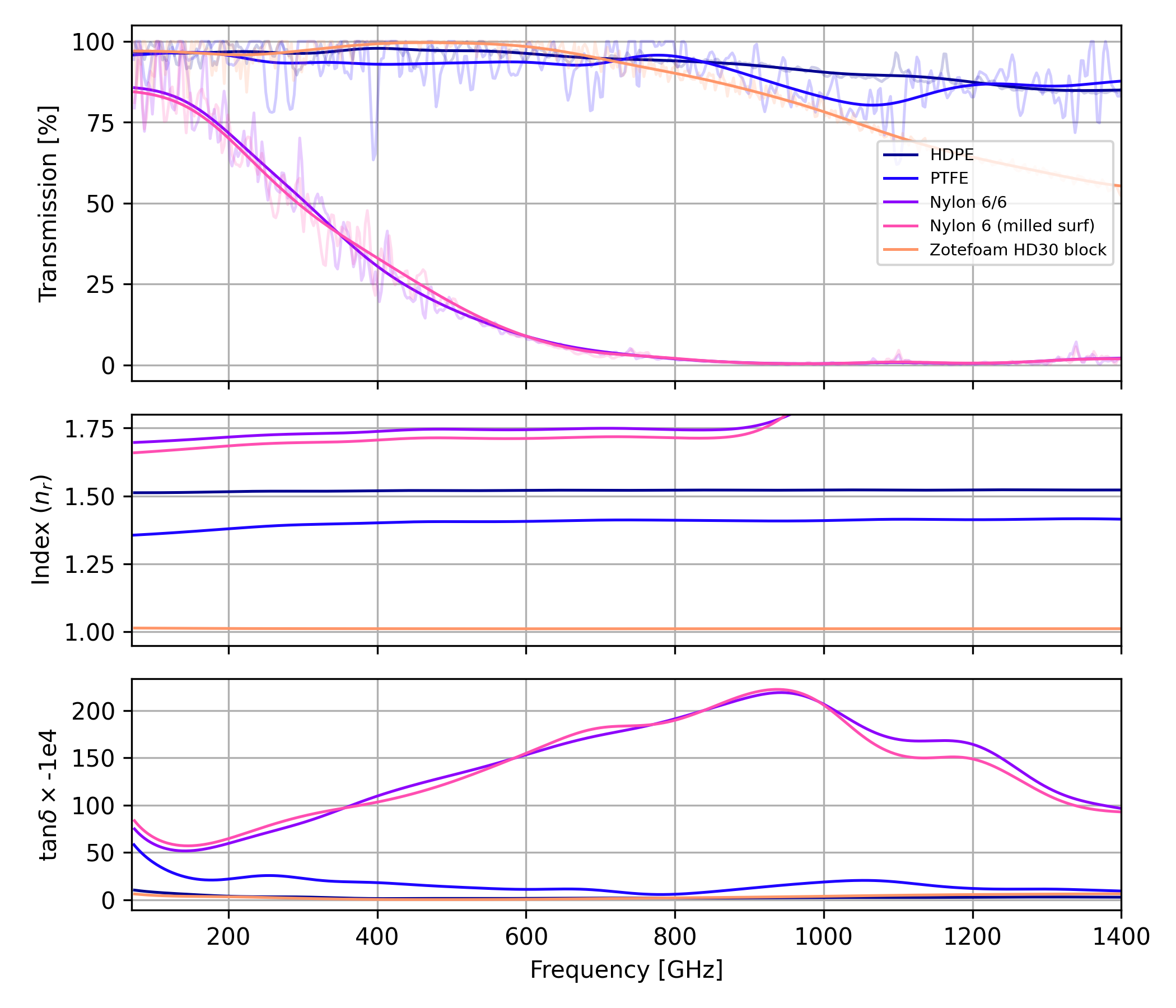}
    \caption{[Top] Transmission over frequency from TeraScan. Lighter alpha lines are raw transmission, dark lines are processed transmission. [Middle] Index of refraction calculated from the measurement. [Bottom] Calculated tan$\delta$ from the measurement.}
    \label{fig:topt_trans}
\end{figure}

We report the TeraScan transmission measurements for these same materials in Figure \ref{fig:topt_trans}. It is important to note that we did not measure the same HD30 stack as was used for the scattering scans: the stack was too thick to fit between the collimating mirrors in the TeraScan, so a block with  roughly the equivalent integrated thickness was used instead (the HD30 block is \SI{25}{\milli\meter} thick, the integrated stack thickness is \SI{27}{\milli\meter}). The data are processed as described in Section \ref{sec:terascan}. The real part of the index of refraction is estimated like $n_r = \sqrt{\epsilon_1}$ and the loss tangent is estimated from $\tan\delta = \epsilon_2/\epsilon_1$. The index of refraction is remarkably stable across the entire frequency range for most of the materials. The only exception to the calculated index of refraction stability is are the two nylon formulations, though the sharp increase in index past \SI{900}{\giga\hertz} is more likely due to the extremely low transmission at higher frequencies leading to complications in the fitting procedure. The loss tangent is similarly stable for the low loss materials, but highly variable in the PTFE and the nylons, which may be expected for materials that have been used as a lossy infrared filter.

We take the average of the optical parameters across the measured range and report the values in Table \ref{tab:parameters}. Most materials (the HDPE, PTFE, and Zotefoam HD30) had relatively stable index of refraction and tan$\delta$ across the entire frequency range, so we report the average from 70--1400 GHz. The nylon parameters, however, vary significantly across the measured range, particularly once the transmission is completely cut off: therefore we only take an average from 70--900 GHz for the nylon samples. These optical parameters for these materials are close to those reported in Lamb, 1996\cite{Lamb1996}: the only major discrepancy is that there does appear to be a difference in index of refraction between the two grades of nylon, with nylon 6/6 having an index closer to the reported value.

\section{Conclusions}\label{sec:conclusion}

\begin{table}[t]
    \centering

    \begin{tabular}{c c c c}
        Material & Avg Index of Refraction ($n_r$) & Avg tan$\delta$ [$\times10^{-4}$] & Freq Range\\
        \hline
        HDPE & 1.52 & 2.5 & Full (70-\SI{1400}{\giga\hertz}) \\
        PTFE & 1.40 & 15.5 & Full (70-\SI{1400}{\giga\hertz}) \\
        Nylon 6/6 & 1.74 & 129 & Partial (70-\SI{900}{\giga\hertz}) \\
        Nylon 6 & 1.70 & 131 & Partial (70-\SI{900}{\giga\hertz}) \\
        HD30 block & 1.01 & 2.9 & Full (70-\SI{1400}{\giga\hertz})
    \end{tabular}
        \caption{Average values of optical parameters for the transmission measurements shown in Figure \ref{fig:topt_trans}. Averages for the nylon samples were taken over a smaller frequency range. Note that these are average values taken over a large frequency range and are meant to be representative of the order of magnitude of the frequency variable parameters reported in the figure.}
    \label{tab:parameters}
\end{table}

We report the empirically measured scattered power from a variety of polymers used in millimeter and sub-millimeter astronomical instruments. Bulk polymers show essentially no excess in scattered power, unless the sample has significant surface structure. Polymer foams like Zotefoam HD30 and LD24 do scatter a significant amount of power out to high angles, likely due to Mie scattering off the relatively large internal cells.

We also report the estimated index of refraction and loss tangent of these polymers out to 1400 GHz in Figure \ref{fig:topt_trans} and Table \ref{tab:parameters}. The material properties are found to be similar to those reported elsewhere in literature\cite{Lamb1996,Elwood2024}.

In future work we will expand our measurements out to more complex materials, such as layered manufactured windows and metal mesh filters. We are eager to explore the potential impact that tolerances or internal structure within anti-reflection coats on scattered power. Additionally, constraining optical parameters of these materials down to the temperatures they typically operate at is paramount: we suspect, for example, that the loss of nylon drops significantly when cold.

\subsection{Next Steps}

We plan to rework the sample holder system to accommodate larger samples: the system described in this work is limited by the carrying capacity of the MECA-500 robot, which is 500 grams. We will adapt a large linear stage to carry the sample holder to potentially allow automation of the normalization measurements. The transmit head will also be placed on an x-y linear stage to enable automation of the focus process: this will enable quick swaps of the frequency adapter heads, enabling near continuous scattering measurements from 70-\SI{330}{\giga\hertz}. 

The TeraScan can also be set up on a rotation stage, enabling 270$^{\circ}$ slice of scattered power to be measured from 70-\SI{1400}{\giga\hertz} \cite{Gaganpreet2024}. We plan to measure these samples in this re-optimized rotation stage apparatus.

\acknowledgments 
Funded by the European Union (ERC, CMBeam, 101040169). Views and opinions expressed are however those
of the author(s) only and do not necessarily reflect those of the European Union or the European Research
Council Executive Agency. Neither the European Union nor the granting authority can be held responsible for
them. We acknowledge support from The Icelandic Research Fund (Grant number: 2410656-051), the Swedish
Research Council (Reg. no. 2019-03959), and the Swedish National Space Agency (SNSA/Rymdstyrelsen).

\bibliography{main} 
\bibliographystyle{spiebib} 

\end{document}